\documentclass{pasj01}
\draft
\Received{$\langle$reception date$\rangle$}
\Accepted{$\langle$acception date$\rangle$}
\Published{$\langle$publication date$\rangle$}
\usepackage{ccaption}
\usepackage{color}
\usepackage{lscape}
\usepackage {threeparttable}

\begin{document}

\title{Massive star formation in the Carina nebula complex and Gum 31 -- I. the Carina nebula complex}
\author{Shinji FUJITA$^{1, 2*}$}
\author{Hidetoshi SANO$^{1}$}%
\author{Rei ENOKIYA$^{1}$}%
\author{Katsuhiro HAYASHI$^{1}$}%
\author{Mikito KOHNO$^{1}$}%
\author{Kisetsu TSUGE$^{1}$}%
\author{Kengo TACHIHARA$^{1}$}%
\author{Atsushi Nishimura$^{2}$}%
\author{Akio OHAMA$^{1}$}%
\author{Yumiko YAMANE$^{1}$}%
\author{Takahiro OHNO$^{1}$}%
\author{Rin YAMADA$^{1}$}%
\author{Yasuo FUKUI$^{1}$}%

\altaffiltext{1}{Department of Astrophysics, Nagoya University, Furo-cho, Chikusa-ku, Nagoya, Aichi, Japan 464-8602}
\altaffiltext{2}{Department of Physical Science, Graduate School of Science, Osaka Prefecture University 1-1 Gakuen-cho, Naka-ku, Sakai, Osaka 599-8531, Japan}

\email{fujita.shinji@a.phys.nagoya-u.ac.jp}



\maketitle

\begin{abstract}
Herein, we present results from observations of the $^{12}$CO ($J$=1--0), $^{13}$CO ($J$=1--0), and $^{12}$CO ($J$=2--1) emission lines toward the Carina nebula complex (CNC) obtained with the Mopra and NANTEN2 telescopes.
We focused on massive-star-forming regions associated with the CNC including the three star clusters Tr~14, Tr~15, and Tr~16, and the isolated WR-star HD~92740.
We found that the molecular clouds in the CNC are separated into mainly four clouds at velocities $-27$, $-20$, $-14$, and $-8$\,km\,s$^{-1}$.
Their masses are $0.7\times 10^{4}$\,$M_{\odot}$, $5.0\times 10^{4}$\,$M_{\odot}$, $1.6\times 10^{4}$\,$M_{\odot}$, and $0.7\times 10^{4}$\,$M_{\odot}$, respectively. 
Most are likely associated with the star clusters, because of their high $^{12}$CO ($J$=2--1)/$^{12}$CO ($J$=1--0) intensity ratios and their correspondence to the {\it Spitzer} 8$\mu$m distributions.
In addition, these clouds show the observational signatures of cloud--cloud collisions.
In particular, there is a V-shaped structure in the position--velocity diagram and a complementary spatial distribution between the $-20$\,km\,s$^{-1}$ cloud and the $-14$\,km\,s$^{-1}$ cloud.  
\textcolor{black}{Furthermore, we found that SiO emission, which is a tracer of a shocked molecular gas, is enhanced between the colliding clouds by using ALMA archive data. }
Based on these observational signatures, we propose a scenario wherein the formation of massive stars in the clusters was triggered by a collision between the two clouds. 
By using the path length of the collision and the assumed velocity separation, we estimate the timescale of the collision to be $\sim$\,1\,Myr.
This is comparable to the ages of the clusters estimated in previous studies. 
\end{abstract}

\section{Introduction}
\subsection{Massive star formation}
Massive stars are influential in the galactic environment, because they release heavy elements and large amounts of energy in the form of ultraviolet radiation, stellar winds, outflows, and supernova explosions. 
It is therefore of fundamental importance to understand the formation mechanisms of massive stars, and considerable efforts have been made to date (e.g., \cite{1987ApJ...319..850W}, \cite{2007ARA&A..45..481Z}, \cite{2014prpl.conf..149T}). 
Most remarkably, recent observational studies have increasingly revealed the importance of cloud--cloud collisions (CCCs). 
The recent successful development of observational tools for identifying cloud--cloud collisions, --i.e., the complementary spatial distributions and bridging features between the colliding clouds--has enabled us to create a firm basis for studying CCC (e.g., \cite{Fuk18}).
When two clouds collide, one burrows into the other owing to momentum conservation (\cite{2015MNRAS.450...10H}).
As a simple example, if a collision takes place head-on between two clouds of different sizes, a cavity will be formed in the larger one through this process, and the larger cloud will apear as a ring-like structure on the plane of the sky, unless the observer's viewing angle is perfectly perpendicular to the collision axis.
As the size of the cavity corresponds to that of the smaller cloud, an observer with a viewing angle parallel to the collision axis sees a complementary distribution between the smaller cloud and the ring-like structure.
The bridging feature is relatively weak CO emission at intermediate velocities between the two colliding clouds, which are separated in the position--velocity ($p$--$v$) diagram.
When one cloud collided with another, a dense compressed and turbulent layer is formed at the collision interface. 
If one observes a snapshot of this collision at a viewing angle parallel to the collision axis, two velocity peaks separated by intermediate-velocity emission with lower intensity is seen in the $p$--$v$ diagram. 
The turbulent gas that creates the bridging feature can be replenished as long as the collision continues.
Many observational studies have reported detections of such complementary spatial distributions and bridging features in CCC regions (e.g., \cite{2018PASJ...70S..49E, 2017arXiv170605664F, 2019PASJ..tmp...46F, 2019ApJ...872...49F, 2014ApJ...780...36F, 2016ApJ...820...26F, 2017PASJ...69L...5F, 2009ApJ...696L.115F, 2019ApJ...886...14F, 2018PASJ...70S..48H, 2018PASJ...70S..50K, 2010ApJ...709..975O, 2018PASJ...70S..43S, 2019ApJ...886...15T, 2015ApJ...806....7T, 2017ApJ...835..142T, 2017ApJ...840..111T, 2018PASJ..tmp..121T, 2018PASJ...70S..51T, 2019PASJ..tmp...50T}). 
The $p$--$v$ diagram of the colliding clouds exhibits a V-shaped structure that depends upon the conditions such as the viewing angle and cloud sizes. 

\subsection{The Carina nebula complex}
The Carina nebula complex (CNC) and Gum 31 are located in the Carina spiral arm (e.g., \cite{2014ApJS..215....1V}), and the CNC is one of the most active massive-star-forming regions in the Milky Way. 
Approximately 140 massive OB-stars (\cite{Ale16}) and more than 1400 young stellar objects (\cite{Pov11}) have been identified in the CNC.
The distance to the CNC, $\sim$2.3\,kpc, has been measured accurately by near-infrared spectroscopy observations (e.g., \cite{All93, Smi06a}). 
\textcolor{black}{We adopt this value in this paper.}
The number of O-stars in the CNC is comparable to that in other active massive-star-forming regions in the Milky Way such as W43 and W51 (e.g., \cite{Blu99, Oku00}), but the CNC is two or three times nearer than those regions (e.g., \cite{Zha14, Sat10}).
Therefore, the CNC offers unique observational advantages for studies of massive-star formation. 
To date, observations of CO emission covering the entire region of the CNC have been conducted. 
The $^{12}$CO ($J$=1--0) map obtained with the Mopra Telescope by \citet{2000PhDT.........2B} achievied a 45\,$''$ resolution.
Subsequently, maps of 3--4 arcmin resolution have been obtained for the $^{12}$CO ($J$=1--0), $^{13}$CO ($J$=1--0), and C$^{18}$O ($J$=1--0) emission lines using the NANTEN telescope (\cite{2005ApJ...634..476Y}). 
Most recently, \citet{Reb16} reported column densities for the molecular clouds associated with the CNC and Gum 31 by using the high-resolution $^{12}$CO and $^{13}$CO ($J$=1--0) emission data obtained with Mopra.
They found regional variations in the column densities obtained from the fraction of the mass recovered from the CO emission lines relative to the total mass traced by the dust emission.
However, the velocity structures of the associated clouds have not previously been analyzed in detail.

\begin{figure}[htbp]
  \begin{center}
  \includegraphics[width=14cm]{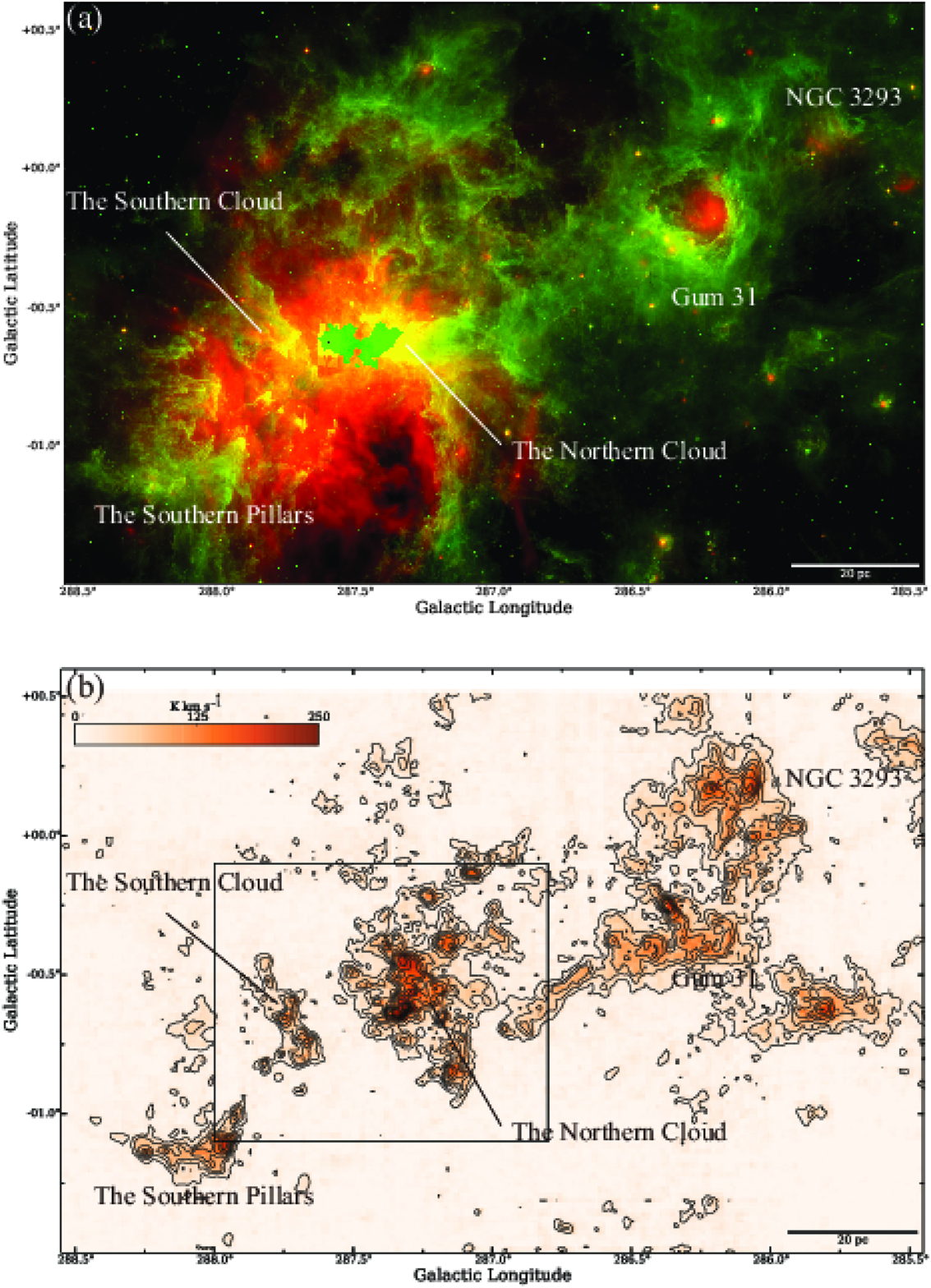}
  \end{center}
  \caption{(a) Composite color image of the WISE 22\,$\mu$m (red) and {\it Spitzer}/GLIMPSE 8\,$\mu$m (green) emissions around the CNC and Gum~31.  (b) Integrated-intensity map of the $^{12}$CO ($J$=1--0) emission taken from the Mopra archive (\cite{Reb16, Reb17}) between the velocities of -35 and 0\,km\,s$^{-1}$. The black contours are plotted every 25\,K\,km\,s$^{-1}$ from 25\,K\,km\,s$^{-1}$ ($\sim 10\sigma$). The black rectangle outlines the area shown in Figure\,\ref{fig:integs}.}\label{fig:RGB}
\end{figure}

Figure\,\ref{fig:RGB}(a) shows a composite color image of the WISE 22 $\mu$m (red) and {\it Spitzer}/GLIMPSE 8\,$\mu$m (green) emissions around the CNC and Gum 31. 
The 22\,$\mu$m emission, which traces mainly hot dust from H{\sc ii} regions, extends over $\sim$ 50\,pc through the CNC, and the data near the center of the CNC [$(l,\,b)=(287\fdg5,\,-0\fdg6)$] are saturated. 
On the other hand, Gum 31 is relatively a small and resembles to the {\it Spitzer} bubbles (\cite{2006ApJ...649..759C, 2007ApJ...670..428C}), although the radius is somewhat larger ($\sim$10\,pc) than is typical for those bubbles. 
In this paper, we focus on the area outlined by the black rectangle in Figure\,\ref{fig:RGB}(b), which includes the star clusters Trumpler~16 (Tr~16), Trumpler~14 (Tr~14), and Trumpler~15 (Tr~15). 
We will focus on Gum 31 in forthcoming paper (paper II).

\section{Dataset}

\subsection{Mopra}\label{sec:dat_m}

We used the $^{12}$CO and $^{13}$CO ($J$=1--0) archived datasets obtained with the Mopra 22-m telescope (\cite{Bur13, Bra15, Reb16, Reb17}).
The beam size and velocity resolution are $\sim$35$''$ and $\sim$0.09\,km\,s$^{-1}$, respectively.
The typical rms noise levels for the $^{12}$CO and $^{13}$CO ($J$=1--0) datasets at a velocity grid of 0.09\,km\,s$^{-1}$ are $\sim$4.0\,K and $\sim$1.9\,K per channel on the $T_{\rm mb}$ scale.
Details of the observations, calibration, and data reduction are summarized in \citet{Bur13}. 

\subsection{NANTEN2}\label{sec:dat_n}

The $^{12}$CO ($J$=2--1) observations were made using the NANTEN2 4m millimeter/sub-millimeter telescope at Atacama, Chile in 2015 October. 
The half-power beam width of NANTEN2 at $\sim$230\,GHz corresponds to $\sim$90$''$, and we adopted the on-the-fly (OTF) mapping mode with Nyquist sampling. 
The $^{12}$CO ($J$=2--1) emissions were obtained with the 4\,K cooled Nb SIS DSB mixer receiver, with a typical system noise temperature including the atmosphere of 300\,K--440\,K. 
A Fourier digital spectrometer installed on the backend of the beam transmission system provides data resolved into 16384 channels at 1\,GHz bandwidth.
We smoothed them to a velocity grid of 0.5\,km\,s$^{-1}$. 
We summed up two orthogonal scan maps (the Galactic longitude-scan and the Galactic latitude-scan maps) with 15$'$$\times$15$'$ binning to reduce scanning effects.
A typical uncertainty in the intensity is $\sim$20\%. 
We determined the scaling factor used to convert the intensities to absolute intensities from pixel-by-pixel comparisons between the $^{12}$CO ($J$=2--1) maps obtained with NANTEN2 ($T_{\rm a}^{*}$) and with the 1.85\,m telescope ($T_{\rm mb}$) toward Orion B (Onishi et al. 2013; Nishimura et al. 2015). 
The typical rms noise level for the $^{12}$CO ($J$=2--1) data at the velocity grid of 0.5\,km\,s$^{-1}$ is $\sim$0.3\,K per channel on the $T_{\rm mb}$ scale.

\subsection{\textcolor{black}{ALMA}}\label{sec:dat_a}
\textcolor{black}{
We used the SiO ($v=0$, $J$=2--1) at 86.84696\,GHz and H$^{13}$CO$^+$ ($J$=1--0) at 86.75428\,GHz of the ALMA 7\,m array archived datasets. 
The observations were conducted during the ALMA Cycle 4 under the project code of 2016.1.01609.S (\cite{Reb20}).
The beam size and velocity resolution are $\sim$15$''$ and $\sim$0.5\,km\,s$^{-1}$, respectively, at the final data cube.
The details of the observations are described in \citet{Reb20}. 
}


\section{Results}\label{sec:res}

\subsection{Large-scale gas distributions}\label{sec:res_large}

\begin{figure}[htbp]
  \begin{center}
  \includegraphics[width=17cm]{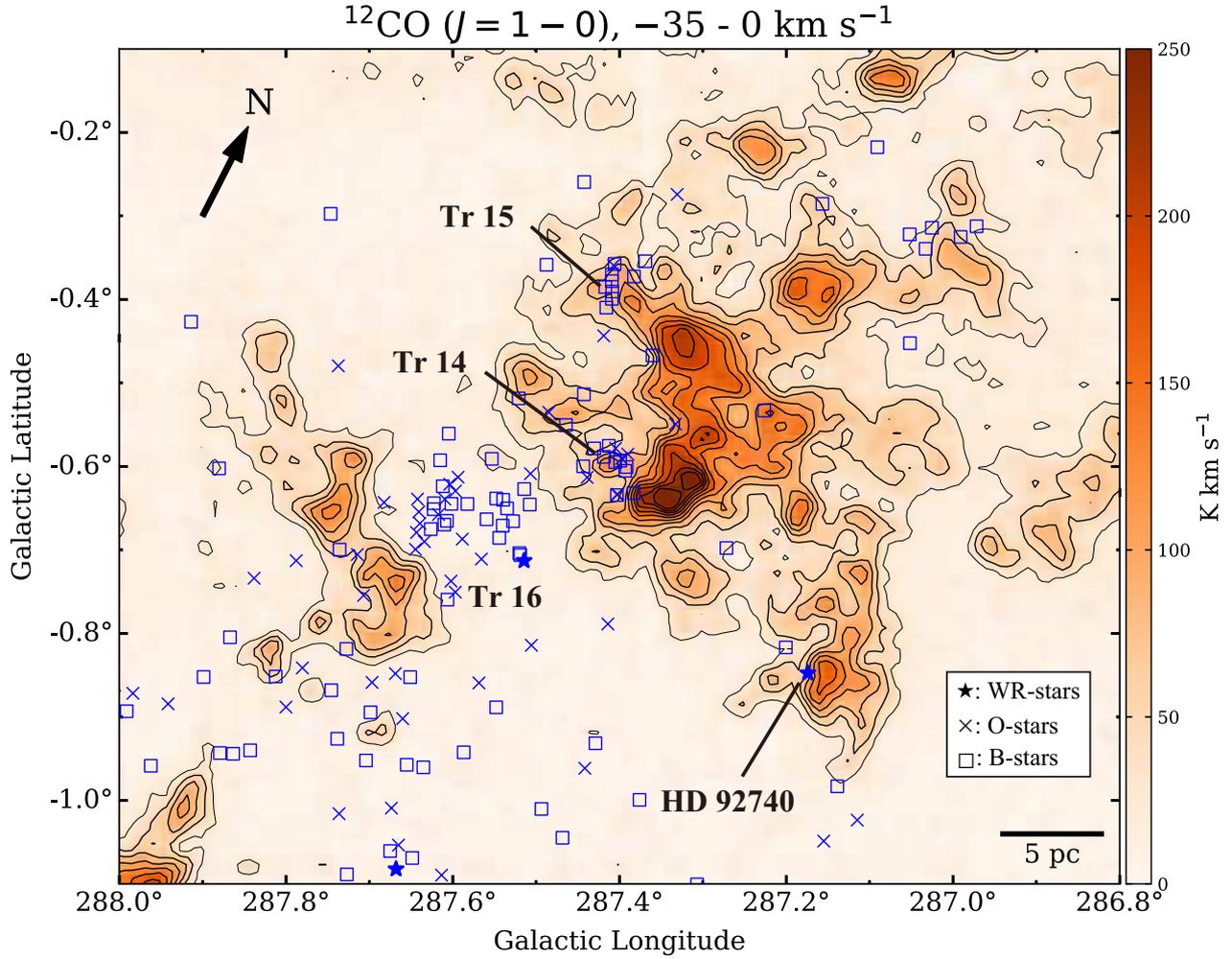}
  \end{center}
  \caption{A closeup figure of Figure\,\ref{fig:RGB}(b). The star symbols represent WR-stars (\cite{Ham06}). The crosses and squares represent O-stars and B-stars, respectively (\cite{Ale16}).}\label{fig:integs}
\end{figure}

Figure\,\ref{fig:RGB}(b) shows the integrated-intensity distributions of the $^{12}$CO ($J$=1--0) emission obtained with the Mopra telescope in the velocity range from $-35$ to $0$\,km\,s$^{-1}$. 
Figure\,\ref{fig:integs} shows a closeup version that includes the massive stars listed by \citet{Ham06} and \citet{Ale16}. 
The detected clouds can be separated mainly into a northern part and a southern part. 
\citet{Reb16} called these the Northern Cloud and the Southern Cloud, respectively. 
The Northern Cloud may be associated with the star clusters Tr~14 and Tr~15. 
The isolated WR-star HD~92740 is also associated with the Northern Cloud.
On the other hand, the center of Tr~16 is located between the Northern Cloud and the Southern Cloud.

\subsection{Velocity structures and CO ratio}\label{sec:vel}

\begin{figure}[htbp]
  \begin{center}
  \includegraphics[width=17cm]{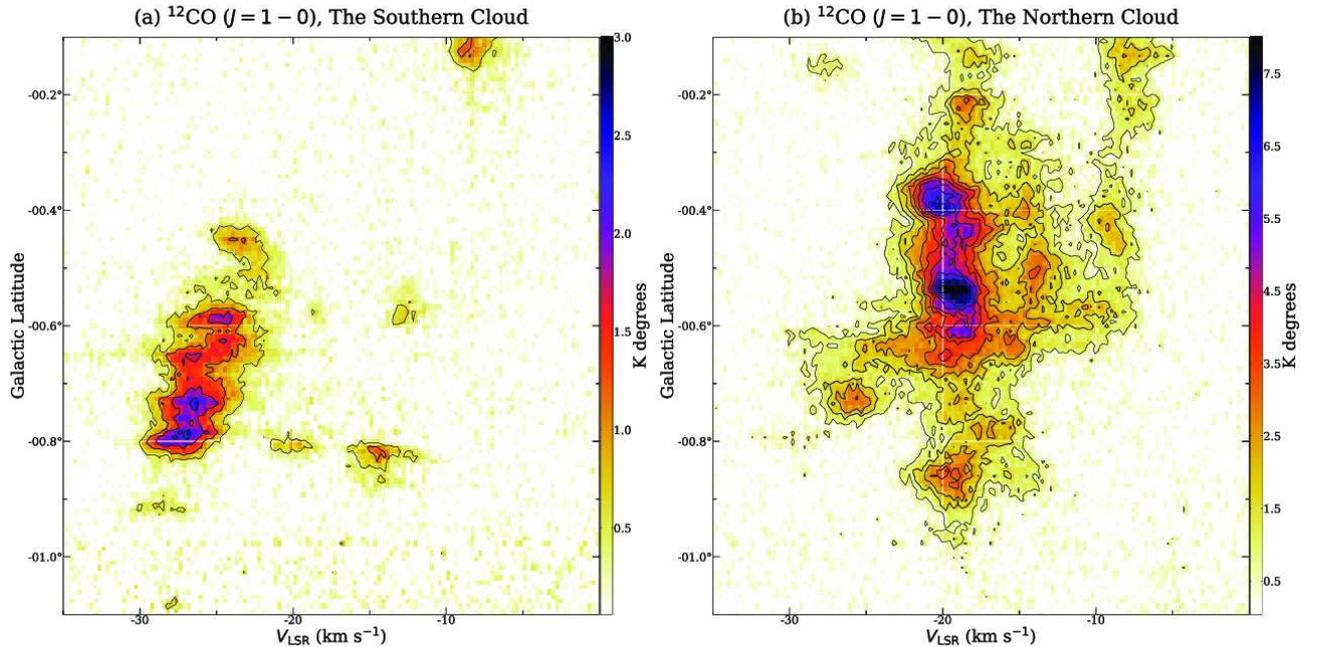}
  \end{center}
  \caption{(a) Velocity--Galactic latitude ($v$--$b$) diagram of the $^{12}$CO ($J$=1--0) emissions from the Southern Cloud integrated from 287\fdg88 to 287\fdg58. The black contours are plotted at every 0.56\,K\,degree from 0.56\,K\,degree ($\sim 10\sigma$). (b) Velocity--Galactic latitude ($v$--$b$) diagram of the $^{12}$CO ($J$=1--0) emission from the Northern Cloud integrated from 287\fdg55 to 286\fdg95. The black contours are plotted every 0.78\,K\,degree from 0.78\,K\,degree ($\sim 10\sigma$). }\label{fig:vbs}
\end{figure}

Figure\,\ref{fig:vbs} shows the $v$--$b$ diagrams of the $^{12}$CO ($J$=1--0) emission toward (a) the Southern Cloud and (b) the Northern Cloud, respectively. 
The main component of the Southern Cloud has velocities of $-30$ to $-20$\,km\,s$^{-1}$, and relatively weak emissions were also detected at $\sim -20$\,km\,s$^{-1}$ and $\sim -14$\,km\,s$^{-1}$ at $b=\sim -0\fdg080$ and $-0\fdg055$. 
In contrast, the velocity structure of the Northern Cloud consists of several components ranging from $\sim -30$\,km\,s$^{-1}$ to $\sim -5$\,km\,s$^{-1}$, as shown in Figure\,\ref{fig:vbs}(b). 
We identified four clouds separated in velocity and centered at -27, -20, -14, and -8\,km\,s$^{-1}$ (hereinafter called the $-27$\,km\,s$^{-1}$ cloud, the $-20$\,km\,s$^{-1}$ cloud, the $-14$\,km\,s$^{-1}$ cloud, and the $-8$\,km\,s$^{-1}$ cloud, respectively).
Clear self-absorption features are not found anywhere in the region.

Figures\,\ref{fig:IR_integ} (a1), (b1), (c1), and (d1) show the {\it Spitzer}/GLIMPSE 8$\mu$m intensity (green color) and the $^{12}$CO ($J$=1--0) integrated-intensity for the -27, -20, -14, and -8\,km\,s$^{-1}$ cloud, respectively. 
Figures\,\ref{fig:IR_integ} (a2), (b2), (c2), and (d2) show the $^{12}$CO ($J$=1--0) integrated-intensity (contours) and the integrated-intensity ratios $^{12}$CO ($J$=2--1)/$^{12}$CO ($J$=1--0) \textcolor{black}{(color scale)} (hereinafter denoted by $R^{12}_{2-1/1-0}$) for the $-27$, $-20$, $-14$, and $-8$\,km\,s$^{-1}$ clouds, respectively.
\textcolor{black}{The velocity ranges for each are summarized in Table\,\ref{tab:cd}.}

\textcolor{black}{In Figure\,\ref{fig:IR_integ} (a1), the edge of the $-27$\,km\,s$^{-1}$ cloud corresponds to the 8$\mu$m emissions in the Southern Cloud region. 
The 8$\mu$m emission is dominated by polycyclic aromatic hydrocarbon (PAH) emission, which is caused by irradiation of dust (e.g., \cite{2001ApJ...546..273C}).
Therefore, the $-27$\,km\,s$^{-1}$ cloud in the Southern Cloud region may be interacting with Tr~16. 
The CO intensity ratio between two different rotational transitions reflects the kinematic temperature and/or the density of the gas. 
In the Northern Cloud region, the intensity of the 8$\mu$m emission is high and $R^{12}_{2-1/1-0}$ is also high at the edge of the $-27$\,km\,s$^{-1}$ cloud (Figure\,\ref{fig:IR_integ} (a2)), suggesting that the $-27$\,km\,s$^{-1}$ cloud in this region is interacting with Tr~16 and/or Tr~14.
In Figures\,\ref{fig:IR_integ} (b1) and (b2), both the southern edge in the Northern Cloud region and the northern edge in the Southern Cloud region of the $-20$\,km\,s$^{-1}$ cloud correspond to the 8$\mu$m emissions and $R^{12}_{2-1/1-0}$ is high at the edge of the cloud in the Northern Cloud region. 
For the same reasons, the $-20$\,km\,s$^{-1}$ cloud also may be interacting with Tr~16, Tr~14, Tr~15, and HD~92740. }
In Figures\,\ref{fig:IR_integ} (c1) and (c2), the $-14$\,km\,s$^{-1}$ cloud is distributed mainly in the Northern Cloud region. 
The $-14$\,km\,s$^{-1}$ cloud may be interacting with Tr~14, Tr~15, and HD~92740 for the same reasons, but it is not clear whether or not it is also interacting with Tr~16.
In Figures\,\ref{fig:IR_integ} (d1) and (d2), the $-8$\,km\,s$^{-1}$ cloud is distributed only in the Northern Cloud region. 
The emissions in this velocity range are clearly related to Tr~14, but not to Tr~15, because $R^{12}_{2-1/1-0}$ is low ($\sim$0.4).

From a local thermodynamic equilibrium analysis of the $^{12}$CO ($J$=1--0) and $^{13}$CO ($J$=1--0) emission data (\cite{Reb16, Reb17}), we estimated the column densities for the five velocity clouds. 
For the Northern Cloud, we estimate the maximum column densities ($N({\rm H_2})$) to be $2.6\pm0.5 \times 10^{22}\,{\rm cm^{-2}}$, $4.5\pm0.9 \times 10^{22}\,{\rm cm^{-2}}$, $2.0\pm0.4 \times 10^{22}\,{\rm cm^{-2}}$, and $0.6\pm0.1 \times 10^{22}\,{\rm cm^{-2}}$, respectively. 
We estimate maximum $N({\rm H_2})$ for the Southern Cloud to be $4.1\pm0.8 \times 10^{22}\,{\rm cm^{-2}}$. 
The estimated errors are due mainly to the calibration errors of 20\% in the CO dataset.
In this derivation, we assumed that the $^{12}$CO ($J$=1--0) emission lines are optically thick, and we derived the excitation temperatures ($T_{\rm ex}$) from the $^{12}$CO ($J$=1--0) peak brightness temperatures for each pixel (the derived $T_{\rm ex}$ is typically 10\,K--40\,K). 
We adopted an abundance ratio of [$^{12}$CO]/[$^{13}$CO]$=77$ (\cite{1994ARA&A..32..191W}) and a fractional $^{12}$CO abundance of $X$($^{12}$CO) = [$^{12}$CO]/[H$_2$]$=10^{-4}$ (\cite{1982ApJ...262..590F, 1984ApJS...56..231L}), which yeild $X$($^{13}$CO) = [$^{13}$CO]/[H$_2$]$=1.3\times10^{-6}$. 
For the Northern Cloud, we estimated the molecular masses to be $0.7\pm0.1 \times 10^{4}\,M_{\odot}$, $5.0\pm1.0 \times 10^{4}\,M_{\odot}$, $1.6\pm0.3 \times 10^{4}\,M_{\odot}$, and $0.7\pm0.1 \times 10^{4}\,M_{\odot}$, respectively, and we estimate the molecular mass of the Southern Cloud to be $1.1\pm0.2 \times 10^{4}\,M_{\odot}$. 
These parameters are summarized in Table\,\ref{tab:cd}.

\begin{figure}[htbp]
  \begin{center}
  \includegraphics[width=16cm]{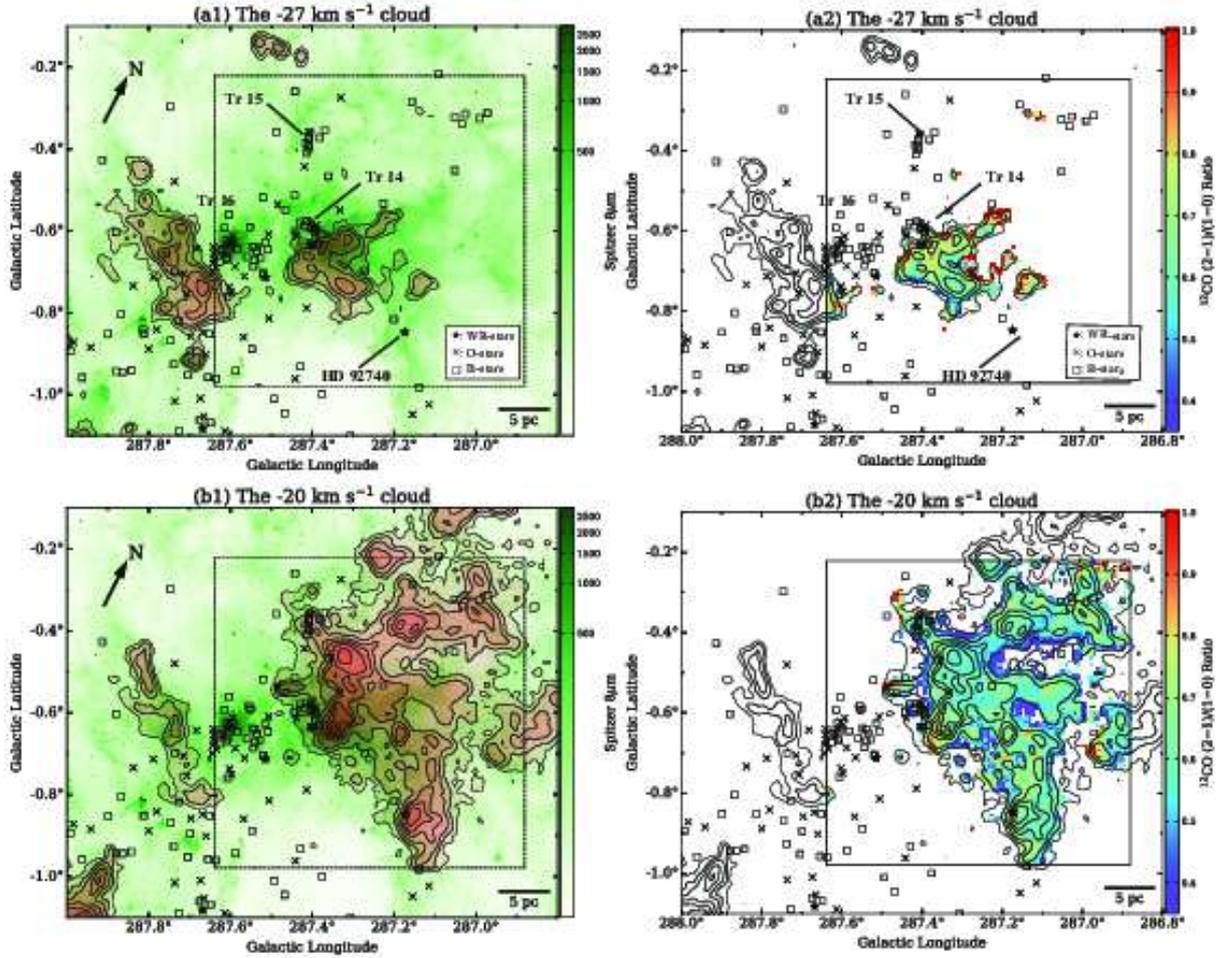}
  \end{center}
  \caption{(a1) The green image shows the intensity of the {\it Spitzer}/GLIMPSE 8$\mu$m emissions. The red filled contours show the integrated-intensity of the $^{12}$CO ($J$=1--0) emissions from the $-27$\,km\,s$^{-1}$ cloud. The contours are plotted at 8, 16, 32, 64, 96, and 128\,K\,km\,s$^{-1}$. The symbols are the same as Figure\,{\ref{fig:integs}}. (a2) Map of the integrated-intensity ratio $^{12}$CO ($J$=2--1)/$^{12}$CO ($J$=1--0) ($R^{12}_{2-1/1-0}$) for the $-27$\,km\,s$^{-1}$ cloud. The black contours are the same as in (a1). The large black square indicates the area of the $^{12}$CO ($J$=2--1) data. The other figures are the same as in (a1) and (a2), but for the $-20$\,km\,s$^{-1}$ cloud, the $-14$\,km\,s$^{-1}$ cloud, and the $-8$\,km\,s$^{-1}$ cloud, respectively. (continued)}\label{fig:IR_integ}
\end{figure}

\begin{figure}[htbp]
  \begin{center}
  \includegraphics[width=16cm]{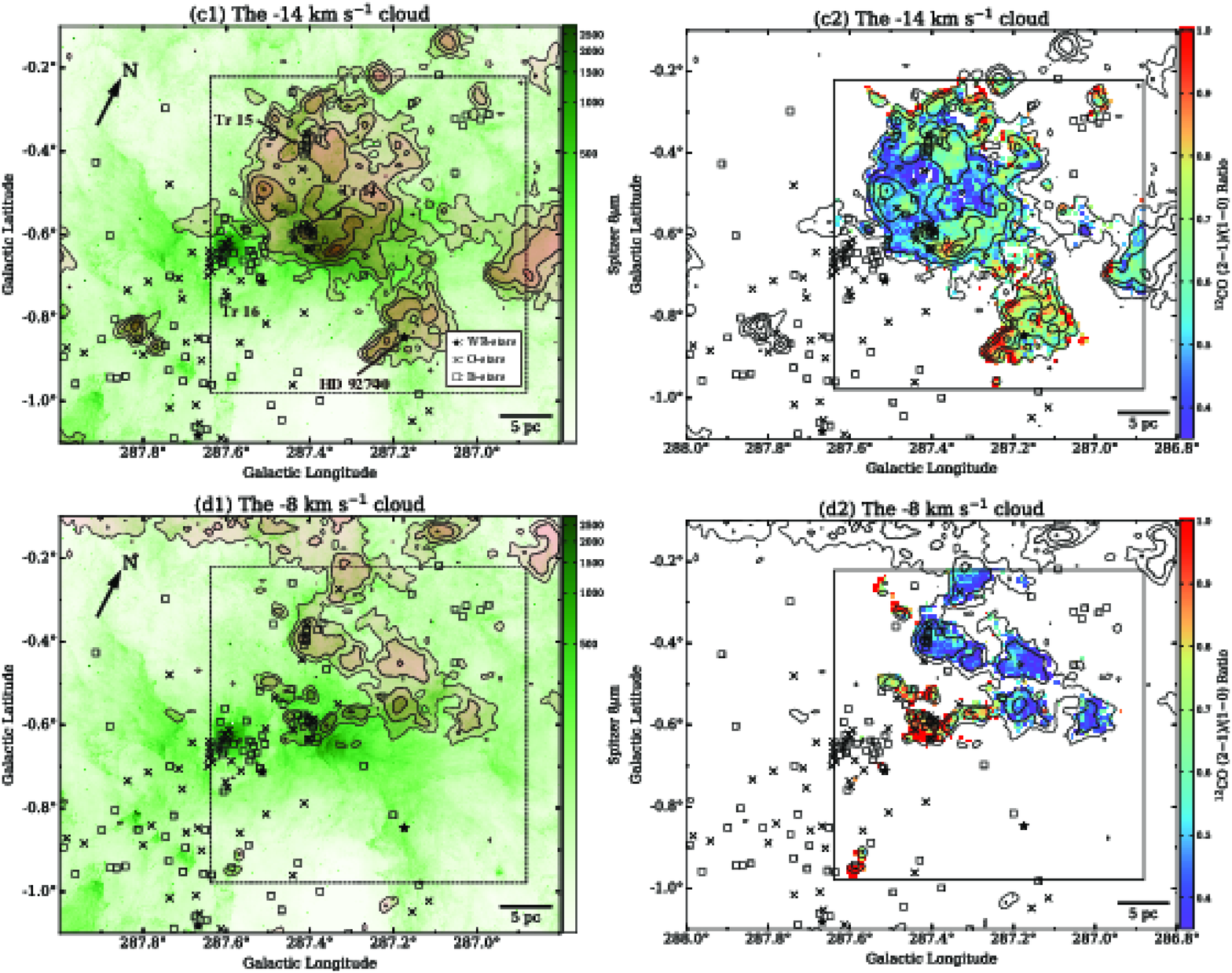}
  \end{center}
  \contcaption{(continued)}
\end{figure}

 \begin{table}[h]
   \tbl{Molecular clouds in the CNC}
   {%
    \begin{tabular}{c|c|c|c|c}
    \hline \hline
    Cloud name & Velocity range & Peak position &$N_{\rm max}({\rm H_2})$  & Mass \\
       & [km\,s$^{-1}$], [km\,s$^{-1}$] & $(l,b)$\,[$^{\circ}$] & [${\rm cm^{-2}}$] & [$M_{\odot}$] \\
    (1) & (2) & (3) & (4) & (5) \\   
       \hline \hline
       The Northern Cloud & & & \\
        The $-27$\,km\,s$^{-1}$ cloud & from $-30$ to $-24$ & $(287.34, -0.64)$ & $2.6\pm0.5 \times 10^{22}$ & $0.7\pm0.1 \times 10^{4}$\\
        The $-20$\,km\,s$^{-1}$ cloud & from $-23$ to $-17$ & $(287.15, -0.85)$ & $4.5\pm0.9 \times 10^{22}$ & $5.0\pm1.0 \times 10^{4}$\\
        The $-14$\,km\,s$^{-1}$ cloud & from $-17$ to $-11$ & $(287.51, -0.49)$ & $2.0\pm0.4 \times 10^{22}$ & $1.6\pm0.3 \times 10^{4}$\\
        The $-8$\,km\,s$^{-1}$ cloud &  from $-11$ to $-5$ & $(287.32, -0.23)$ & $0.6\pm0.1 \times 10^{22}$ & $0.7\pm0.1 \times 10^{4}$\\ \hline
        The Southern Cloud & & & &\\
         & from $-30$ to $-20$ & $(287.67, -0.74)$ & $4.1\pm0.8 \times 10^{22}$ & $1.1\pm0.2 \times 10^{4}$\\
       \hline
    \end{tabular}} \label{tab:cd}
\begin{tabnote}
(1) Name of the cloud. 
(2) Velocity range of the cloud. 
(3) Peak position of the H$_2$ column density. 
(4) Maximum H$_2$ column density. 
(5) Mass of the cloud. 
\end{tabnote}
\end{table}

\clearpage

\section{Discussion}\label{sec:Dis}

\subsection{The molecular clouds associated with Tr~14, Tr~15, Tr~16, and HD~92740, and cloud--cloud collisions}\label{sec:mca}
In previous sections we found that CO emissions are divided into several velocity components.
The area of the present study includes three large clusters and one isolated WR-star; Tr~14, Tr~15, Tr~16, and HD~92740.
In order to investigate the star-formation history in the CNC, we here consider the molecular clouds associated with each cluster.

\subsubsection{Tr~14}\label{sec:Tr14}

\begin{figure}[htbp]
  \begin{center}
  \includegraphics[width=16cm]{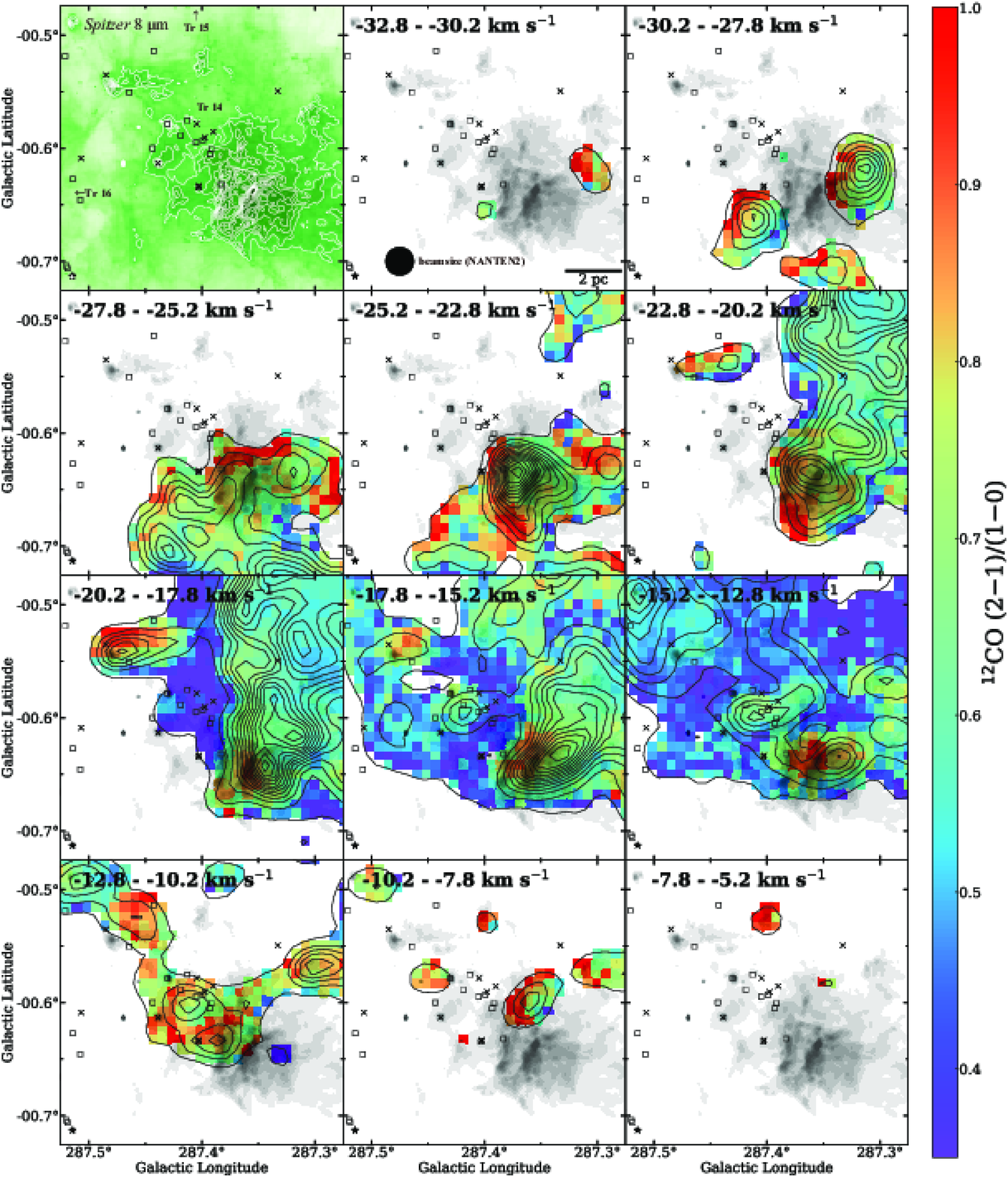}
  \end{center}
  \caption{Velocity channel maps of the integrated-intensity ratio ($^{12}$CO ($J$=2--1)/$^{12}$CO ($J$=1--0)) toward Tr~14. The top-left panel and the black contours in the other panels show the {\it Spitzer} 8\,$\mu$m emissions. The 8\,$\mu$m contours are plotted at every 100\,M\,Jy\,str$^{-1}$ from 300\,M\,Jy\,str$^{-1}$. The CO integration range for each panel is given in the top-left corner of the panel. The CO contours are plotted every 6\,K\,km\,s$^{-1}$ from 6\,K\,km\,s$^{-1}$. The symbols are the same as in Figure\,\ref{fig:integs}. \textcolor{black}{The black square in the top-left panel indicate the area of the ALMA data shown in Figure\,\ref{fig:chmap_12CO_SiO}.}}\label{fig:Tr14ch}
\end{figure}

Toward Tr~14, many velocity components overlap along the line of sight. 
Figure\,\ref{fig:Tr14ch} shows a velocity-channel map of the CO integrated-intensity ratio $R^{12}_{2-1/1-0}$ overlaied on the {\it Spitzer} 8\,$\mu$m emissions toward Tr~14. 
In the velocity range $-30.2$ to $-20.2$\,km\,s$^{-1}$, CO emissions have been detected only in the region where the intensity of 8\,$\mu$m emissions are high and the $R^{12}_{2-1/1-0}$ is higher ($\sim 1.0$) at the edges of the clouds, indicating that these molecular clouds are probably interacting with the H{\sc ii} region of Tr~14. 
Such characteristics of molecular clouds are observed in other H{\sc ii} regions as well; e.g., W~51 (\cite{2019PASJ..tmp...46F}), which is one of the most active massive star forming region.
In the velocity range \textcolor{black}{$-20.2$ to $-12.8$\,km\,s$^{-1}$, $R^{12}_{2-1/1-0}$ is relatively high ($\sim 0.8$) at the edge of the cloud in the western side of Tr~14 ($l,\,b=\sim 287\fdg37,\,\sim -0\fdg65$), although the ratio is low where the CO intensity is low}.
\textcolor{black}{These results indicate that both the $-20$\,km\,s$^{-1}$ cloud and the $-14$\,km\,s$^{-1}$ cloud are also interacting with the H{\sc ii} region of Tr~14, while there may be non-interacting diffuse cloud at the velocity range of $-20.2$ to $-12.8$\,km\,s$^{-1}$.}
In the velocity range above \textcolor{black}{$-12.8$\,km\,s$^{-1}$}, $R^{12}_{2-1/1-0}$ is again high ($\sim 1.0$), whereas the CO intensity is relatively low. 
These emissions may include outflows from young massive stars in Tr~14, as discussed by \citet{2005ApJ...634..476Y}. 
We need more detailed observations to identify them.

\begin{figure}[htbp]
  \begin{center}
  \includegraphics[width=12cm]{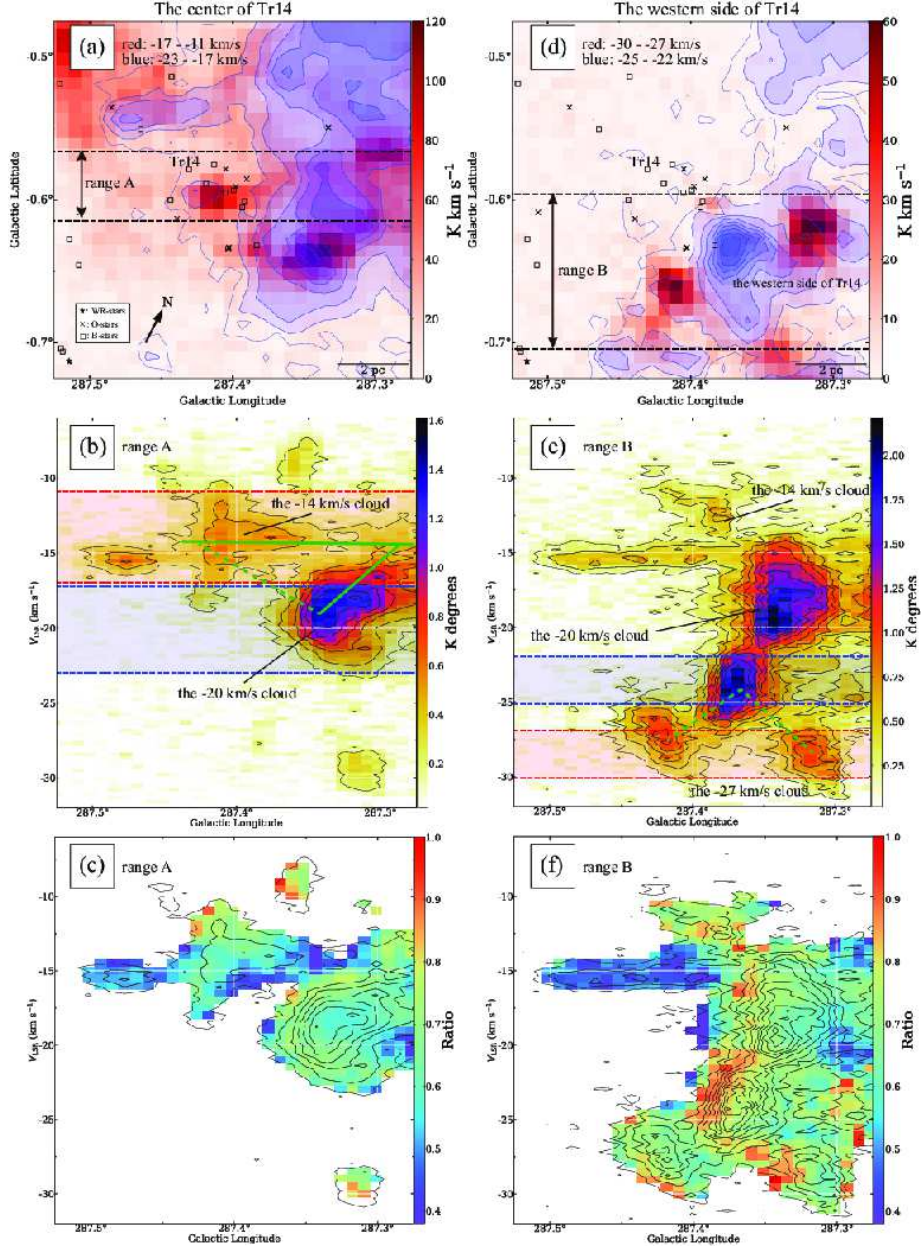}
  \end{center}
  \caption{(a) The red image and blue contours show the integrated-intensity of the $^{12}$CO ($J$=1--0) emissions toward the center of Tr~14. The integrated velocity ranges are shown by the red and blue shading in (b). The contours are plotted at 8, 16, 32, 64, 96, 128, and 160\,K\,km\,s$^{-1}$. The symbols are the same as in Figure\,{\ref{fig:integs}}. (b) Galactic longitude--Velocity ($l-v$) diagram of the $^{12}$CO ($J$=1--0) emissions integrated over the range A shown by the dashed lines in (a). The contours are plotted every 0.2\,K\,degree from 0.2\,K\,degree. (c) Galactic longitude--Velocity ($l-v$) diagram of the intensity ratio $^{12}$CO ($J$=2--1)/$^{12}$CO ($J$=1--0). (d--f) The same as (a--c), but toward the western side of Tr~14.}\label{fig:Tr14lv}
\end{figure}

We next investigate the detailed velocity structures of the molecular clouds around Tr~14. 
Figure\,\ref{fig:Tr14lv}(a) shows the $^{12}$CO ($J$=1--0) integrated-intensity of the $-20$\,km\,s$^{-1}$ cloud and the $-14$\,km\,s$^{-1}$ cloud toward Tr~14.
Figures\,\ref{fig:Tr14lv}(b) and (c) show the velocity--Galactic latitude ($v$--$b$) diagram of the $^{12}$CO ($J$=1--0) emissions and $R^{12}_{2-1/1-0}$, respectively, toward the center of Tr~14. 
The integration range is indicated by the black dashed lines (range A) in Figure\,\ref{fig:Tr14lv}(a). 
The ratio $R^{12}_{2-1/1-0}$ is slightly high ($\sim \,0.7$) in the $-20$\,km\,s$^{-1}$ cloud and in the higher velocity range of the $-14$\,km\,s$^{-1}$ cloud. 
These results suggest that the two molecular clouds are indeed associated with Tr~14. 
Figure\,\ref{fig:Tr14lv}(d) shows the $^{12}$CO ($J$=1--0) integrated-intensity in the velocity ranges of $-30$ to $-27$\,km\,s$^{-1}$ and $-25$ to $-22$\,km\,s$^{-1}$. 
Figures\,\ref{fig:Tr14lv}(e) and (f) are the same as Figures\,\ref{fig:Tr14lv}(b) and (c), but toward the western side of Tr~14 [range B in Figure\,\ref{fig:Tr14lv}(d)]. 
The ratio $R^{12}_{2-1/1-0}$ of the two velocity components is high ($\sim 1.0$), suggesting that the two are interacting with the H{\sc ii} region. 
We found that the two velocity components exhibit a complementary spatial distribution, which is one of the significant observational signatures of a CCC (e.g., \cite{Fuk18}).
The two can be separated in the $v$--$b$ diagram, indicating that they cannot be explained as a velocity gradient in a molecular cloud. 
In addition, they show reversed V-shaped structures in Figures\,\ref{fig:Tr14lv}(e) and (f). 
Such V-shaped structures in position--velocity diagrams have been observed in other objects (e.g., \cite{2019PASJ..tmp...46F}) as a significant observational signature of a CCC. 

\begin{figure}[htbp]
  \begin{center}
  \includegraphics[width=16cm]{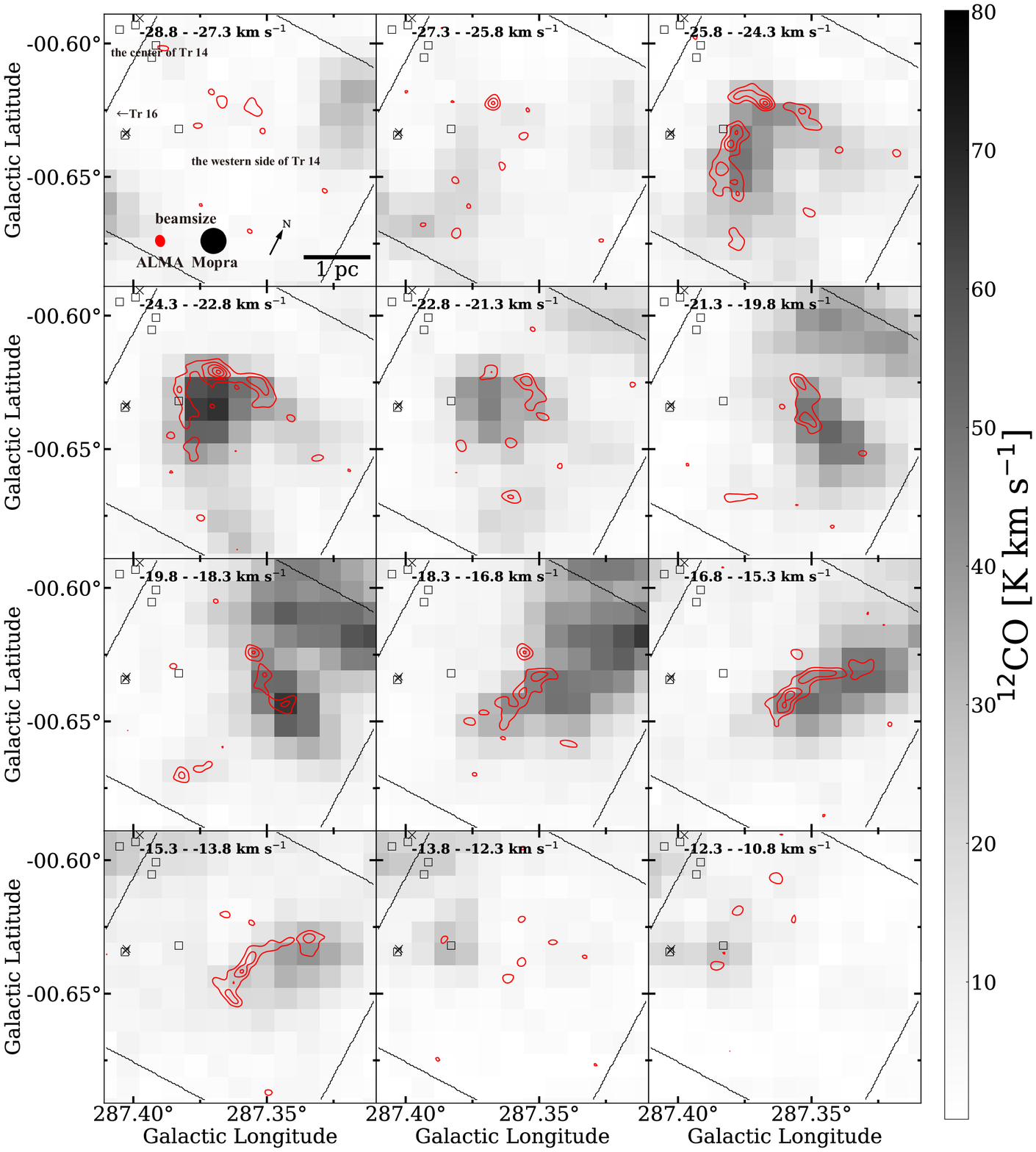}
  \end{center}
  \caption{\textcolor{black}{Velocity channel maps of the integrated-intensity of $^{12}$CO ($J$=1--0) (gray scale) and SiO ($v$=0, $J$=2--1) (red contours) obtained with ALMA toward the western side of Tr~14. The symbols are the same as in Figure\,{\ref{fig:integs}}. The contours are plotted every 0.12\,Jy\,beam$^{-1}$\,km\,s$^{-1}$ from 0.12\,K\,km\,s$^{-1}$. The black lines indicate the area of the ALMA data. }}\label{fig:chmap_12CO_SiO}
\end{figure}

\begin{figure}[htbp]
  \begin{center}
  \includegraphics[width=16cm]{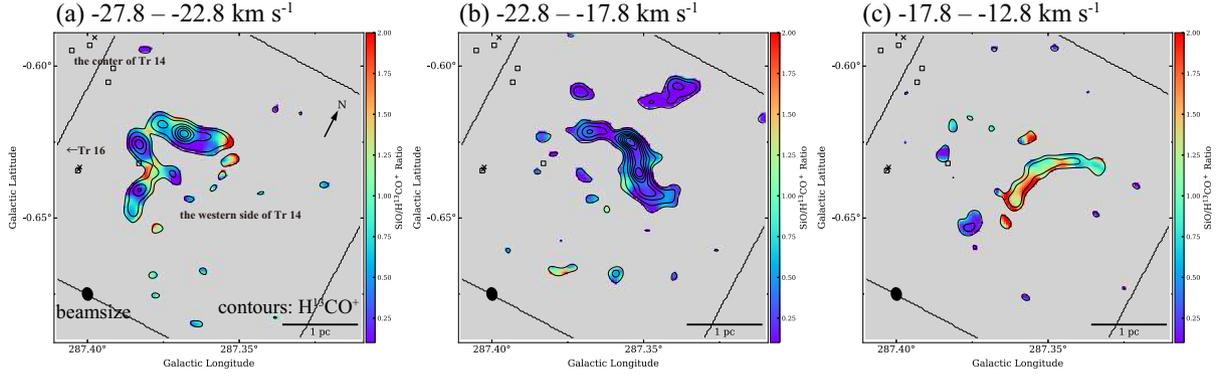}
  \end{center}
  \caption{\textcolor{black}{The integrated-intensity ratio SiO ($v$=0, $J$=2--1)/H$^{13}$CO$^+$ ($J$=1--0) toward the western side of Tr~14. The contours show the integrated-intensity of H$^{13}$CO$^+$ ($J$=1--0). The contours are plotted every 0.2\,Jy\,beam$^{-1}$\,km\,s$^{-1}$ from 0.1\,Jy\,beam$^{-1}$\,km\,s$^{-1}$. }}\label{fig:SiO_H13CO_ratio_integ}
\end{figure}

\begin{figure}[htbp]
  \begin{center}
  \includegraphics[width=16cm]{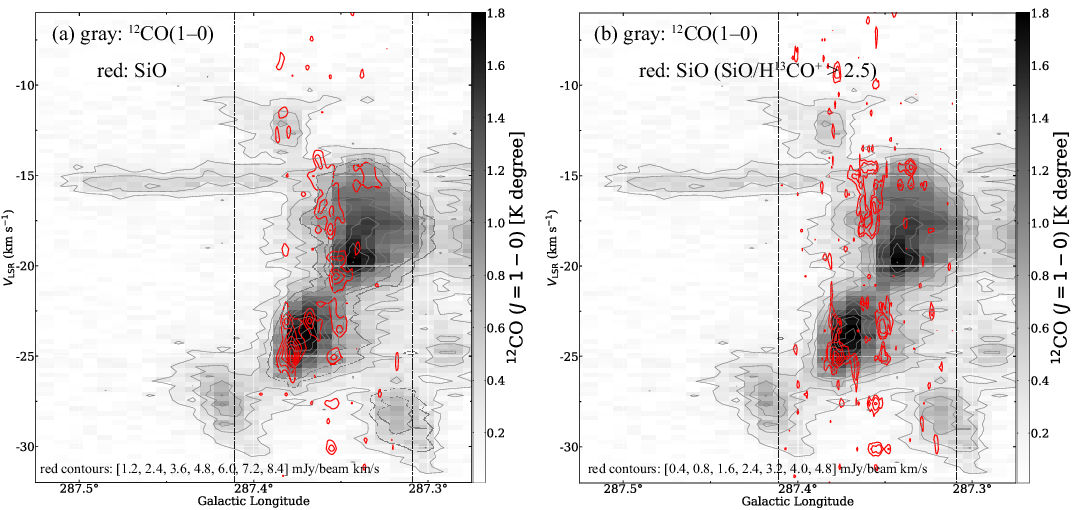}
  \end{center}
  \caption{\textcolor{black}{(a) Galactic longitude--Velocity ($l-v$) diagram of the $^{12}$CO ($J$=1--0) (gray scale and gray contours) and SiO ($v$=0, $J$=2--1) (red contours) integrated from $b=-0.68$ to $b=-0.61$. The gray contours are plotted every 0.2\,K\,degree from 0.2\,K\,degree. The red contours are plotted every 1.2\,mJy\,beam$^{-1}$\,degree ($\sim \,5\sigma$) from 1.2\,mJy\,beam$^{-1}$\,degree. The black vertical dashed-lines indicate the area of the ALMA data. (b) The same as (a), but integrated only the voxels having SiO/H$^{13}$CO$^+$ ratio of higher than 2.5. The red contours are plotted every 1.2\,mJy\,beam$^{-1}$\,degree from 0.6\,mJy\,beam$^{-1}$\,degree.}}\label{fig:lv_12CO_SiO}
\end{figure}

\textcolor{black}{To search other evidence of a CCC in the western side of Tr~14, we next investigate the shocked gas in this region by using the SiO ($v$=0, $J$=2--1) and H$^{13}$CO$^+$ ($J$=1--0) emission data obtained with ALMA [2016.1.01609.S (\cite{Reb20})].
The thermal SiO lines are thought to be a good tracer of hot and shocked gas because the abundance of SiO molecules in a gas phase increase in a high-temperature ($T_{\rm k}\,>\,\sim \,100$\,K) environment (e.g., \cite{1989ApJ...343..201Z}). 
Figure\,\ref{fig:chmap_12CO_SiO} shows the velocity channel maps of the integrated-intensity of $^{12}$CO ($J$=1--0) (gray scale) and SiO ($v$=0, $J$=2--1) (red contours) toward the western side of Tr~14. 
The detected SiO emissions are not masers because they have diffused structures in several parsec.
In the velocity range $-24.3$ to $-22.8$\,km\,s$^{-1}$, SiO emissions are detected with an arc-like structure at the edge of the molecular cloud traced by the $^{12}$CO ($J$=1--0) line. 
These SiO emissions are thought to be tracing PDR induced by the nearby massive stars in Tr~16 and the center Tr~14, as suggested in \cite{Reb20}.
On the other hand, H$^{13}$CO$^+$ ($J$=1--0), whose rest frequency is close to that of SiO ($v$=0, $J$=2--1), is known as a dense gas tracer, and their ratio SiO/H$^{13}$CO$^+$ is often used as a shocked molecular gas tracer (e.g., \cite{2006ApJ...636..261H, 2011A&A...526A..54A, 2015PASJ...67..109T, 2019ApJ...872..121U}).
Figure\,\ref{fig:SiO_H13CO_ratio_integ} shows the integrated-intensity ratio SiO ($v$=0, $J$=2--1)/H$^{13}$CO$^+$ ($J$=1--0) toward the western side of Tr~14. 
Since there is no ratio gradient in the direction of the clusters (Tr~16 and the center of Tr~14), indicating that this ratio is enhanced due to mechanisms other than the radiation from the nearby massive stars.
Figure\,\ref{fig:lv_12CO_SiO}(a) shows the $l-v$ diagram of the $^{12}$CO ($J$=1--0) (gray scale and gray contours) and SiO ($v$=0, $J$=2--1) (red contours) integrated from $b=-0.68$ to $b=-0.61$. 
Figure\,\ref{fig:lv_12CO_SiO}(b) is the same as Figure\,\ref{fig:lv_12CO_SiO}(a), but integrated only the voxels having SiO/H$^{13}$CO$^+$ ratio of higher than 2.5. 
This threshold of the ratio is the same value that was used to identify CCCs in \citet{2019ApJ...872..121U}. 
We can see that the higher ratio gas distributes mainly in three points, $(287\fdg38, -25\,{\rm km s}^{-1})$, $(287\fdg36, -17\,{\rm km s}^{-1})$, and $(287\fdg35, -23\,{\rm km s}^{-1})$. 
In the $l-v$ diagram, the gas in $(287\fdg38, -25\,{\rm km s}^{-1})$ is located between the two cloud, whose velocity ranges are $-30$ to $-27$\,km\,s$^{-1}$ and $-25$ to $-22$\,km\,s$^{-1}$, respectively.  
The ratio enhancement of this region is possible to interpreted as being due to the shock induced by the CCC between the cloud at $-25$ to $-22$\,km\,s$^{-1}$ and $-27$\,km\,s$^{-1}$ cloud, which strongly supports the CCC scenario proposed above.
Similarly, the high ratio gas in $(287\fdg36, -17\,{\rm km s}^{-1})$ suggests a collision between the $-20$\,km\,s$^{-1}$ cloud and the $-14$\,km\,s$^{-1}$ cloud.
The high ratio gas in $(287\fdg35, -23\,{\rm km s}^{-1})$ is detected between the cloud at $-25$ to $-22$\,km\,s$^{-1}$ and $-20$\,km\,s$^{-1}$ cloud. 
This result may indicate a collision between them, but we cannot conclude at this time.
For these reasons, we propose a CCC scenario involving these four clouds as the triggering mechanism for the formation of the massive stars in the western side of Tr~14.
}

\subsubsection{Tr~15}\label{sec:Tr15}
Figure\,\ref{fig:Tr15lv}(a) shows the $^{12}$CO ($J$=1--0) integrated-intensity of the $-20$\,km\,s$^{-1}$ cloud and the $-14$\,km\,s$^{-1}$ cloud toward Tr~15.
One O-star and several B-stars are located at $(l,\,b)=(287\fdg44,\,-0\fdg40)$, where there is both an intensity valley in the $-14$\,km\,s$^{-1}$ cloud and the edge of the $-20$\,km\,s$^{-1}$ cloud. 
For Figures\,\ref{fig:IR_integ}(b2) and (c2), the ratios $R^{12}_{2-1/1-0}$ are high ($\sim 0.8$) both the $-20$\,km\,s$^{-1}$ cloud and the $-14$\,km\,s$^{-1}$ cloud near Tr~15, indicating that the two are interacting with the H{\sc ii} region of Tr~15.
The $-8$\,km\,s$^{-1}$ cloud is also detected in this region, but it is probably not interacting with the H{\sc ii} region because of its low $R^{12}_{2-1/1-0}$ ($\sim 0.4$). 
Figures\,\ref{fig:Tr15lv}(b) and (c) show the $v$--$b$ diagram of the $^{12}$CO ($J$=1--0) emissions and $R^{12}_{2-1/1-0}$, respectively, toward Tr~15. 
We found that the $v$--$b$ diagram shows a V-shaped structure. 
This characteristic in the position--velocity diagram resembles those in the numerical simulation of a CCC by \citet{2014ApJ...792...63T} and \citet{2015MNRAS.450...10H}. 
In this simulation, two molecular clouds of different sizes (one twice as large as the other) collide at a relative velocity of 5\,km\,s$^{-1}$ (see Table\,3 in \citet{Fuk18}).
Figure\,\ref{fig:simu} shows the position--velocity ($p-v$) diagram of an artificial observation of this simulation at the viewing angle of 45$^{\circ}$.
Owing to this viewing angle, the $p-v$ diagram shows an asymmetric feature. 
Similarly to the western side of Tr~14, a CCC between the two clouds may have taken place, triggering massive-star formation in Tr~15.

\begin{figure}[htbp]
  \begin{center}
  \includegraphics[width=8cm]{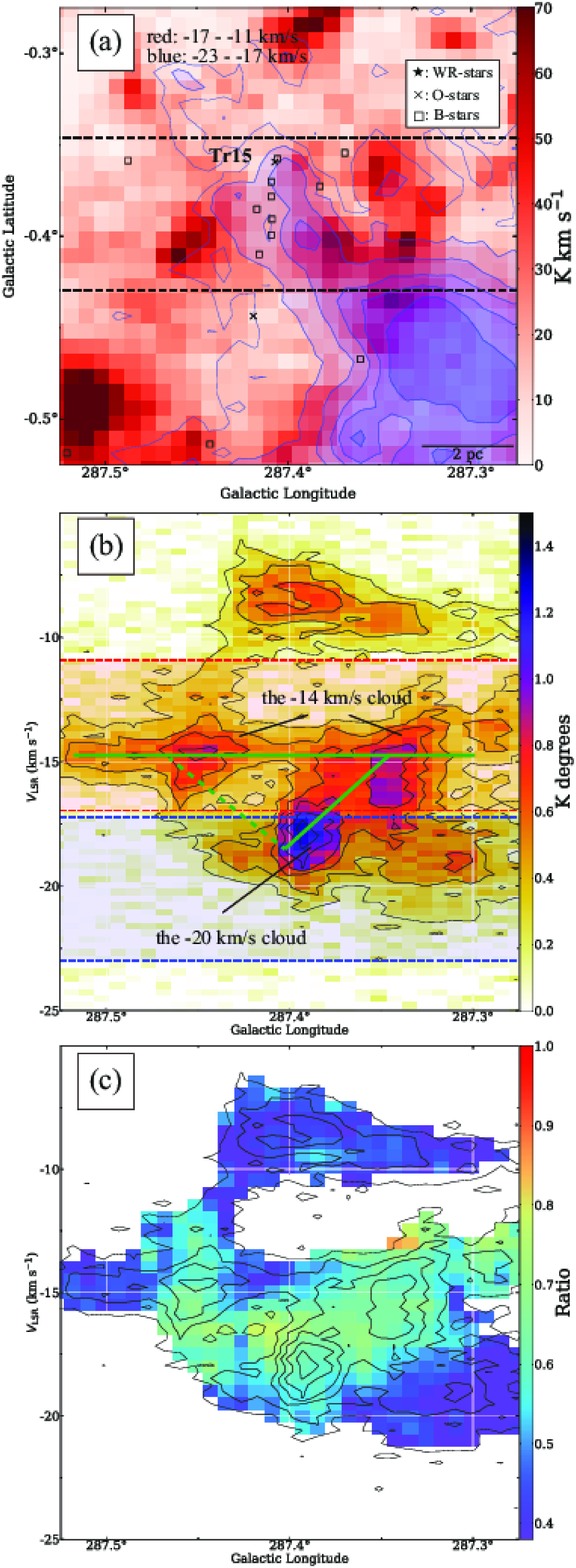}
  \end{center}
  \caption{(a--c) The same as Figure\,\ref{fig:Tr14lv}, but toward Tr~15.}\label{fig:Tr15lv}
\end{figure}

\begin{figure}[htbp]
  \begin{center}
  \includegraphics[width=8cm]{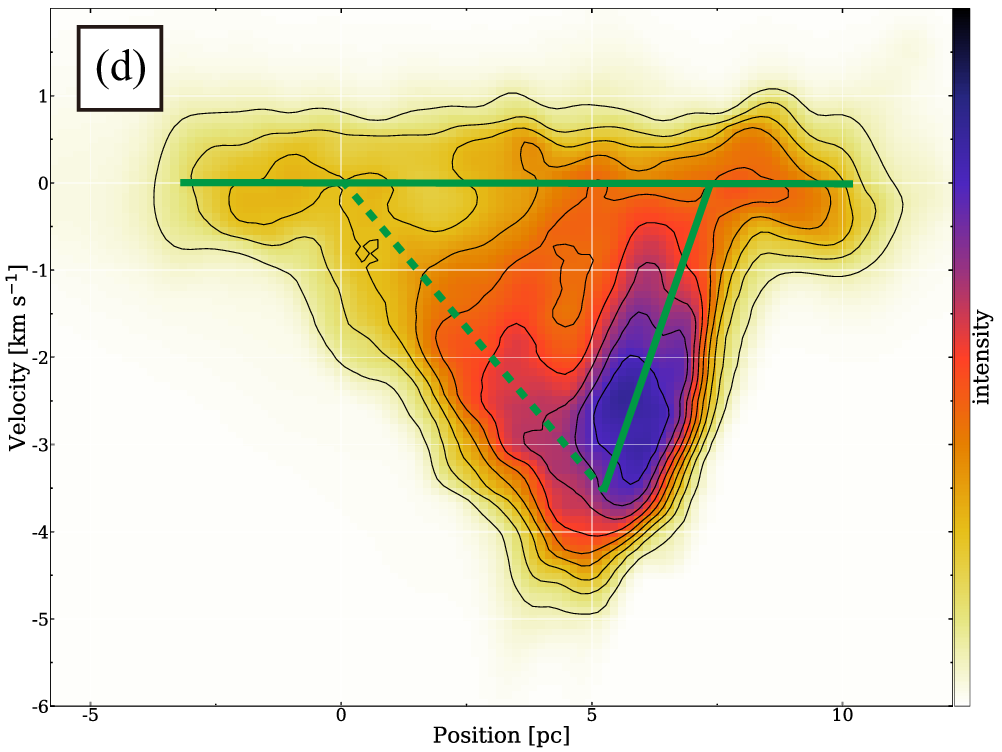}
  \end{center}
  \contcaption{Continued. (d) A position--velocity diagram of the simulation by \citet{2014ApJ...792...63T} and \citet{2015MNRAS.450...10H}, and used in \citet{Fuk18}. The epoch is 1.6\,Myr after the onset of the collision and the viewing angle is 45$^{\circ}$, respectively (see Figure\,4 in \citet{Fuk18}). The integration range is from $-2.5$ to $+2.5$\,pc along an axis perpendicular to the collision axis. }\label{fig:simu}
\end{figure}

\subsubsection{Tr~16}\label{sec:Tr16}
In Figures\,\ref{fig:RGB}(a) and (b), the Southern Cloud is clearly interacting with the H{\sc ii} regions formed by the massive stars in Tr~16 \textcolor{black}{, because the cloud traced by CO corresponds to the distribution of 8$\mu$m emissions}.
The Southern Cloud has a velocity gradient of $\sim 5$\,km\,s$^{-1}$ over the whole area [Figure\,\ref{fig:vbs}(a)]. 
The H{\sc ii} region is already extended by a few tens of parsecs, and weak emissions are detected from the $-20$\,km\,s$^{-1}$ and the $-14$\,km\,s$^{-1}$ cloud at the center of Tr~16. 
In order to investigate the molecular clouds associated with Tr-16, it will be necessary to observe them with both deep sensitivity and high spatial resolution.

\subsubsection{HD~92740}\label{sec:92740}
\begin{figure}[htbp]
  \begin{center}
  \includegraphics[width=8cm]{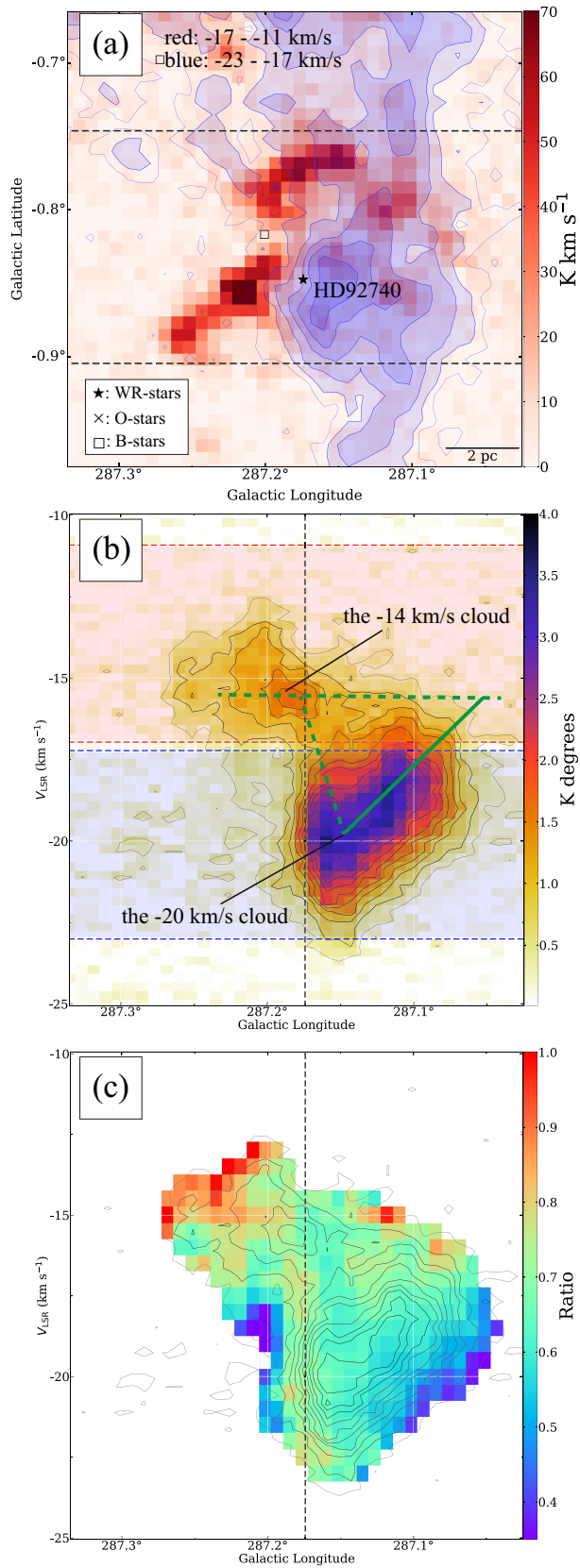}
  \end{center}
  \caption{(a--c) The same as Figure\,\ref{fig:Tr14lv}, but toward HD~92740. The contours in (b) and (c) are plotted every 0.3\,K\,degree from 0.3\,K\,degree. The vertical dashed line in (b) and (c) indicates the position of HD~92740. }\label{fig:92740lv}
\end{figure}

HD~92740 is an isolated WR-star located west of Tr~14.
Figure\,\ref{fig:92740lv}(a) shows the $^{12}$CO ($J$=1--0) integrated-intensity of the $-20$\,km\,s$^{-1}$ cloud and the $-14$\,km\,s$^{-1}$ cloud toward HD~92740.
We found that the two clouds exhibit a complementary distributions also in this region.
HD~92740 is located at the interface between them.
Figures\,\ref{fig:92740lv}(b) and (c) show the $l-v$ diagram of the $^{12}$CO ($J$=1--0) emissions and the $l-v$ diagram of $R^{12}_{2-1/1-0}$, respectively, toward HD~92740. 
In the $l-v$ diagrams, the two clouds are separated into discrete clouds, rather than appearing to be one molecular cloud with a velocity gradient.
\textcolor{black}{For these reasons, it is possible that a CCC between the two clouds took place and triggered the formation of HD~92740 similarly to the western side of Tr~14 and Tr~15. }

\subsection{Cloud--cloud collision scenario in the CNC}\label{sec:ccc}
\begin{figure}[htbp]
  \begin{center}
  \includegraphics[width=17cm]{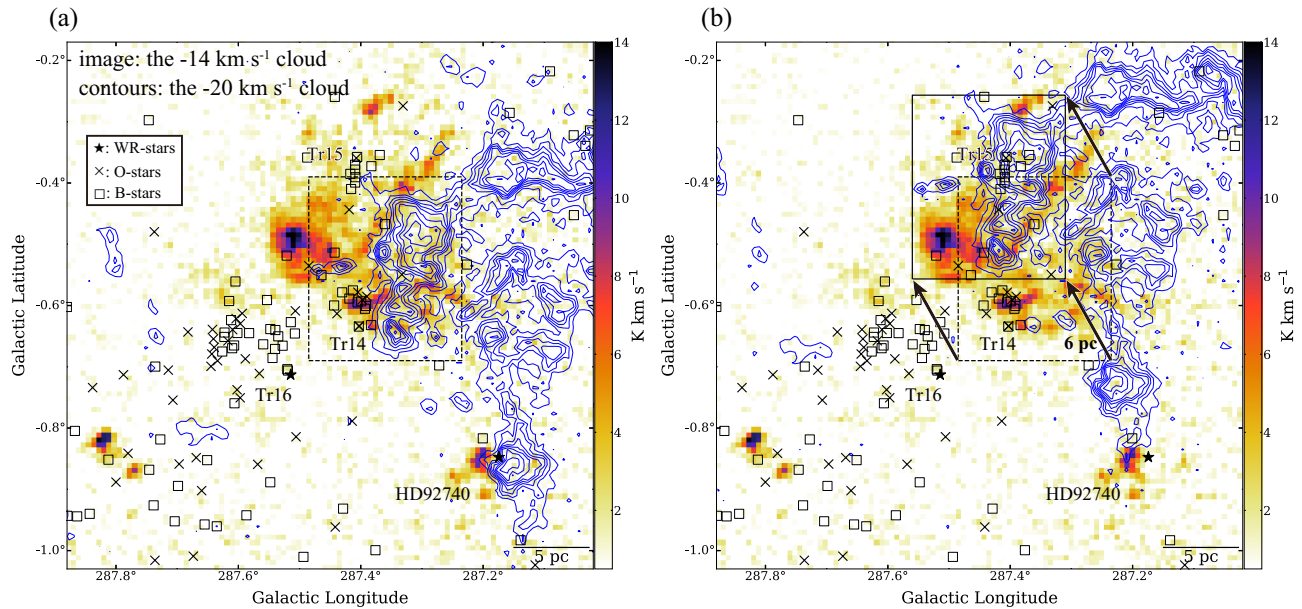}
  \end{center}
  \caption{(a) The color image and blue contours show the integrated-intensity of the $^{12}$CO ($J$=1--0) emissions toward the CNC integrated over $-13.8$ to $-13.3$\,km\,s$^{-1}$ and $-20.9$ to $-20.4$\,km\,s$^{-1}$, respectively. The contours are plotted every 2.2\,K\,km\,s$^{-1}$ (5$\sigma$) from 2.2\,K\,km\,s$^{-1}$. The symbols are the same as in Figure\,{\ref{fig:integs}}. (b) The same as (a), but with the blue contours (the $-20$\,km\,s$^{-1}$ cloud) displaced along the black arrows. }\label{fig:comp}
\end{figure}

As discussed above, we found the signatures of a CCC in three regions: the western side of Tr~14, near Tr~15, and near HD~92740. 
The radial velocities of the pair of colliding clouds in Tr~15 and HD~92740, the $-20$\,km\,s$^{-1}$ cloud and the $-14$\,km\,s$^{-1}$ cloud, are the same, and their $p$--$v$ diagrams are similar to each other. 
Figure\,\ref{fig:comp}(a) shows the $^{12}$CO ($J$=1--0) integrated-intensity distributions of the $-20$\,km\,s$^{-1}$ cloud (contours) and the $-14$\,km\,s$^{-1}$ cloud (image). 
The two clouds show complementary distributions through the entire region of the Northern Cloud, implying that they are colliding over a 10-pc scale.
If so, the $-14$\,km\,s$^{-1}$ cloud may have a cavity of the same size and shape as the $-20$\,km\,s$^{-1}$ clouds. 
This cavity would have been formed at the onset of the collision.

To search for such a cavity, we first quantify the complementarity between the two molecular clouds.
We adopted Spearman's rank correlation coefficient between the integrated-intensity distributions of the two clouds as the index of complementarity.
If the two show a complementary distribution, the correlation coefficient between them is negative; that is, the lower correlation coefficient (anti-correlation) indicates that the complementarity is high.
To calculate the coefficients, we determined the area of the $-20$\,km\,s$^{-1}$ cloud [the contours in Figure\,\ref{fig:comp}(a)] as shown by the black rectangle in Figure\,\ref{fig:comp}(a).
We then define area of the same size for the $-14$\,km\,s$^{-1}$ cloud [the image in Figure\,\ref{fig:comp}(a)] as a function of the X-axis (Galactic longitude) displacement and Y-axis (Galactic latitude) displacement, and we calculated the coefficients between the two intensity distributions. 
In the calculating the correlation coefficients, we removed those pixels with values of $<\,$10\,$\sigma$ for both clouds.
Figure\,\ref{fig:spea} shows the calculated correlation coefficients for each direction of displacement (see Appendix A). 
The black square at ($0$, $0$) indicates the original position (without displacements) of the two integrated-intensity distributions shown in Figure\,\ref{fig:comp}(a). 
We found that the two distributions shows a minimum correlation coefficient ($\sim -0.75$) for a displacement of (X-axis displacement, Y-axis displacement)$=$($-3.0$\,pc, $5.3$\,pc) $\approx$ 6\,pc, indicated by the arrows in Figure\,\ref{fig:comp}(b). 
Figure\,\ref{fig:comp}(b) shows the integrated-intensity distributions of the $-14$\,km\,s$^{-1}$ cloud and the $-20$\,km\,s$^{-1}$ cloud displaced by this value.
This may be the most complementary distribution of the two clouds, indicating that the collision between them started at this relative position where Tr~15 is located.
Accordingly, the massive stars in Tr~15 may have formed early in the collision.

Next, we estimate the timescale of the collision by using the position of the cavity calculated above.
The calculated displacement of $6$\,pc corresponds the cavity length on the plane of the sky . 
Because we cannot measure the velocity difference between the two clouds along the cavity, we tentatively assume that it is the same as the velocity separation along the line-of-sight, 6\,km\,s$^{-1}$. 
Thus, we estimate the timescale of the collision to be roughly $6\,\rm{[pc]}/(6\,\rm{[km\,s}^{-1}])\,=\,\sim \, 1 \, \times \,10^6\,\rm{[yr]}$.
This estimated collision timescale is consistent with the ages of the clusters ($2$--$3$\,Myr for Tr~14 (\cite{2011ApJS..194...10P}) and $6\,\pm \,3$\,Myr for Tr~15 (\cite{1980AJ.....85..708F, 1988PASP..100.1431M, Smi06b})) within a factor of $2$--$3$, although both of these estimates are rough. 

\begin{figure}[htbp]
  \begin{center}
  \includegraphics[width=17cm]{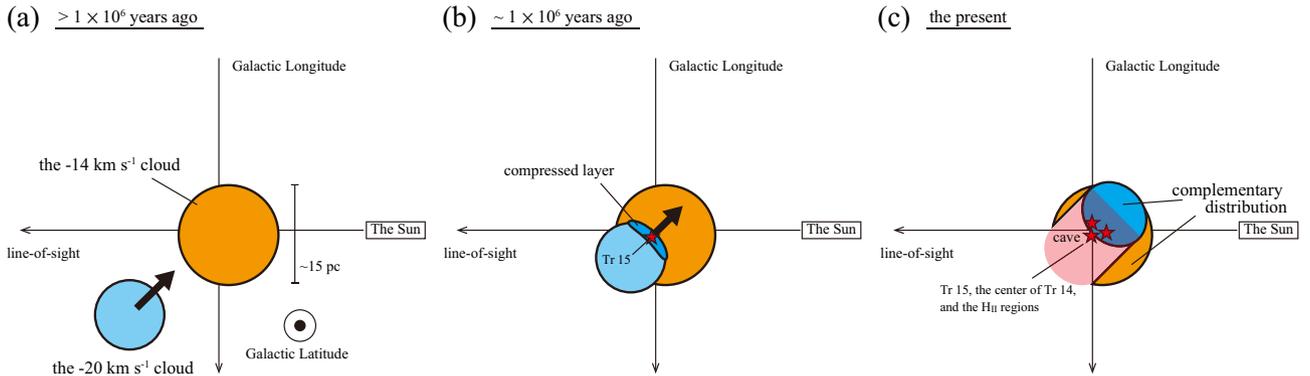}
  \end{center}
  \caption{A sketch diagram of the Northern Cloud as viewed from the Galactic north pole. (a) Before the collision started ($>\,1\, \times \,10^6$ yr ago). Blue and orange represent the $-20$\,km\,s$^{-1}$ cloud and the -14 km\,s$^{-1}$ cloud, respectively. (b) The epoch when the collision has started and a little time has elapsed ($\sim \,1\, \times \,10^6$ yr ago). The deep blue region indicates the compressed layer between the two clouds, where star(s) form. (c) The present. The two clouds show the complementary distributions for observers in the Solar System. }\label{fig:sche}
\end{figure}

In Section\,\ref{sec:Tr14}, we found evidence for a CCC between the molecular clouds with the velocities of $\sim$\,-27\,\,km\,s$^{-1}$ and $\sim$\,-23\,\,km\,s$^{-1}$ in the western side of Tr~14. 
The two clouds are now in a complementary distribution and the number of identified massive stars is small, so probably only a small time has passed since the collision started.
On the other hand, in the center of Tr~14, the $p$--$v$ diagram of the clouds [Figure\,\ref{fig:Tr14lv}(b)] can be interpreted as a CCC between the $-20$\,km\,s$^{-1}$ cloud and the $-14$\,km\,s$^{-1}$ cloud as indicated by the green V-shape as for Tr~15. 
Therefore, the massive stars triggered by these two CCCs may have overlapped along the line of sight or have been mixed into Tr~14. 
Clarifying this situation will require more detailed observations with higher angular resolution.
Figure\,\ref{fig:sche} shows a sketch diagram of this collision scenario between the $-20$\,km\,s$^{-1}$ cloud and the $-14$\,km\,s$^{-1}$ cloud in the Northern Cloud.

In the region near HD~92740, only the one WR-star, HD~92740, formed. 
It is probable that the lack of massive stars in this region may be due to the lower mass of one of the colliding clouds [Figure\,\ref{fig:92740lv}(a)].
Only a few molecular clouds have been associated with the region around Tr~16, perhaps because of strong feedback from the massive stars including $\eta$-Carinae.
It is uncertain whether the formation of the massive stars in Tr~16 is related to the CCC we have suggested.

\begin{figure}[htbp]
  \begin{center}
  \includegraphics[width=10cm]{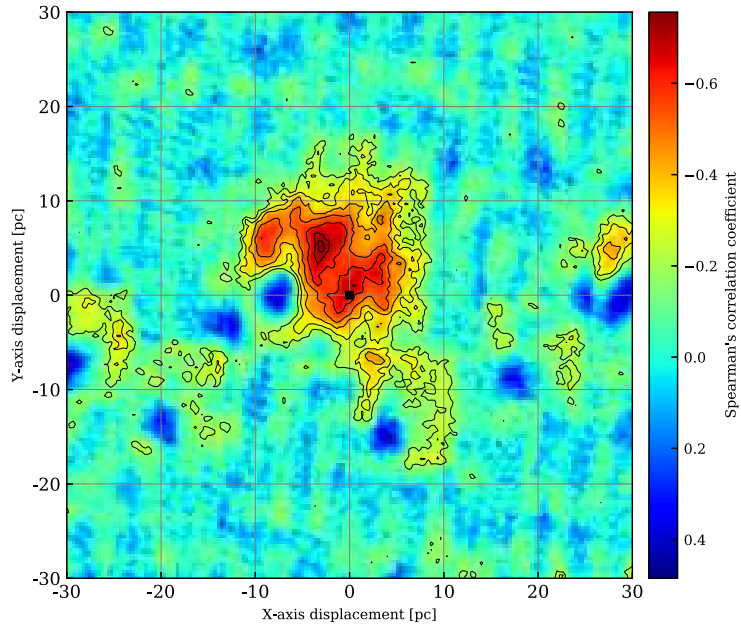}
  \end{center}
  \caption{The distribution of Spearman's correlation coefficient as a function of displacements in the X-axis (Galactic longitude) direction and in the Y-axis (Galactic latitude) direction. The contour levels are -0.74, -0.70, -0.60, -0.50, -0.40, -0.30, and -0.20. }\label{fig:spea}
\end{figure}

\section{Summary}
The main conclusions of the present study are summarized as follows:

\begin{enumerate}
\item We presented analyses of the velocity structures of the molecular clouds in the CNC by using the $^{12}$CO and $^{13}$CO ($J$=1--0) archival datasets obtained with the Mopra telescope and the $^{12}$CO ($J$=2--1) dataset obtained with NANTEN2. 

\item We found that the molecular clouds in the CNC can be separated into four clouds at the velocities $-27$, $-20$, $-14$, and $-8$\,km\,s$^{-1}$. Their masses are $0.7\times 10^{4}$\,$M_{\odot}$, $5.0\times 10^{4}$\,$M_{\odot}$, $1.6\times 10^{4}$\,$M_{\odot}$, and $0.7\times 10^{4}$\,$M_{\odot}$, respectively. Most of them are likely associated with the clusters because of the high $^{12}$CO ($J$=2--1)/$^{12}$CO ($J$=1--0) intensity ratio and the correspondence with the {\it Spitzer} 8$\mu$m distributions. 

\item We found the observational signatures of cloud--cloud collisions in near Tr~14, Tr~15, and HD~92740; namely, a V-shaped structure and a complementary spatial distribution, between the $-20$\,km\,s$^{-1}$ cloud and the $-14$\,km\,s$^{-1}$ cloud. \textcolor{black}{Furthermore, we found that SiO emission, which is a tracer of a shocked molecular gas, is enhanced between the colliding clouds by using ALMA archive data.}

\item We propose a scenario wherein the formation of massive stars in the clusters and the formation of HD~92740 were triggered by a collision between the two clouds. The timescale of the collision is estimated to be $\sim$\,1\,Myr, which is roughly comparable to the ages of the clusters estimated in previous studies. 

\end{enumerate}


\begin{ack}
This study was financially supported by Grants-in-Aid for Scientific Research (KAKENHI) of the Japanese society for the Promotion of Science (JSPS; grant numbers 15K17607 and 17H06740). 
The authors would like to thank the all members of the Mopra, NANTEN2 and ALMA for providing the data. 
Data analysis was carried out by using Astropy (\cite{2013A&A...558A..33A}),  APLpy (\cite{2012ascl.soft08017R}). 
The authors also would like to thank NASA for providing FITS data of the WISE and the {\it Spitzer} Space Telescope. 
\end{ack}

\clearpage

\appendix
\section{H$^{13}$CO$^+$ emission in the western side of Tr~14}
\begin{figure}[htbp]
  \begin{center}
  \includegraphics[width=16cm]{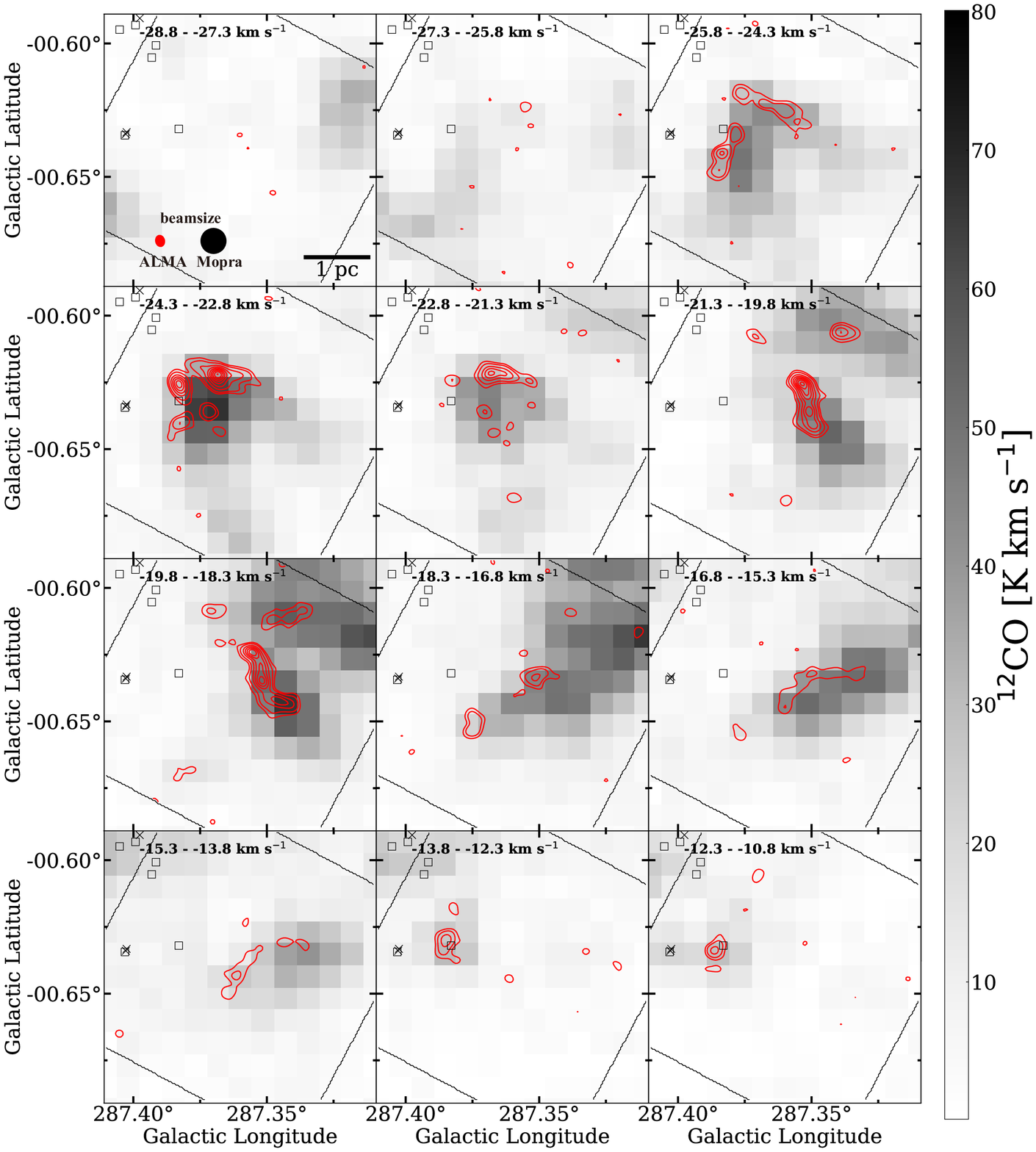}
  \end{center}
  \caption{\textcolor{black}{Velocity channel maps of the integrated-intensity of $^{12}$CO ($J$=1--0) (gray scale) and H$^{13}$CO$^+$ ($J$=1--0) (red contours) obtained with ALMA toward the western side of Tr~14. The symbols are the same as in Figure\,{\ref{fig:integs}}. The contours are plotted every 0.24\,Jy\,beam$^{-1}$\,km\,s$^{-1}$ from 0.12\,K\,km\,s$^{-1}$. The black lines indicate the area of the ALMA data. }}\label{fig:chmap_12CO_H13CO}
\end{figure}

\textcolor{black}{Figure\,\ref{fig:chmap_12CO_H13CO} shows the velocity channel maps of the integrated-intensity of $^{12}$CO ($J$=1--0) (gray scale) obtained with Mopra and H$^{13}$CO$^+$ ($J$=1--0) (red contours) obtained with ALMA toward the western side of Tr~14.}

\section{Complementarity calculation using Spearman's rank correlation coefficient}
If two distributions are similar, the correlation coefficient $r$ ($-1\leq r \leq 1$) between them is a large positive number.
Conversely, if two distributions are complementary to each other, the correlation coefficient between them is low, and perhaps even negative.
We applied this idea to search for the hole (cavity entrance) that the $-20$\,km\,s$^{-1}$ cloud created in the $-14$\,km\,s$^{-1}$ cloud during the collision.
We used Spearman's rank correlation coefficient ($r_S$) to reduce the effects of the extremely high intensity distributions of the molecular clouds.
In statistics, $r_S$ is a nonparametric measure of rank correlation. 
It shows the statistical dependence between the rankings of two variables, namely the integrated intensities in the present case.
For a sample of size $n$ and the integrated intensities $X_i$ (the $-14$\,km\,s$^{-1}$ cloud) and $Y_i$ (the $-20$\,km\,s$^{-1}$ cloud), $r_S$ is computed as 

\begin{equation}
r_S = \frac{{\rm cov}({\rm rg}X,\,{\rm rg}Y)}{\sigma _{{\rm rg}X}\sigma _{{\rm rg}X}}
\end{equation}
where ${\rm rg}X_i$ and ${\rm rg}Y_i$ are the ranks of $X_i$ and $Y_i$, respectively.
The quantities $\sigma$ and cov are the standard deviation and covariance of the ranked variables, respectively.
The correlation coefficient $r_S$ can be computed using the following formula if all $n$ ranks are distinct integers.
\begin{equation}
r_S = 1 - \frac{6\Sigma ({\rm rg}X_i - {\rm rg}Y_i)^2}{n(n^2-1)}. 
\end{equation}

\begin{figure}[htbp]
  \begin{center}
  \includegraphics[width=7cm]{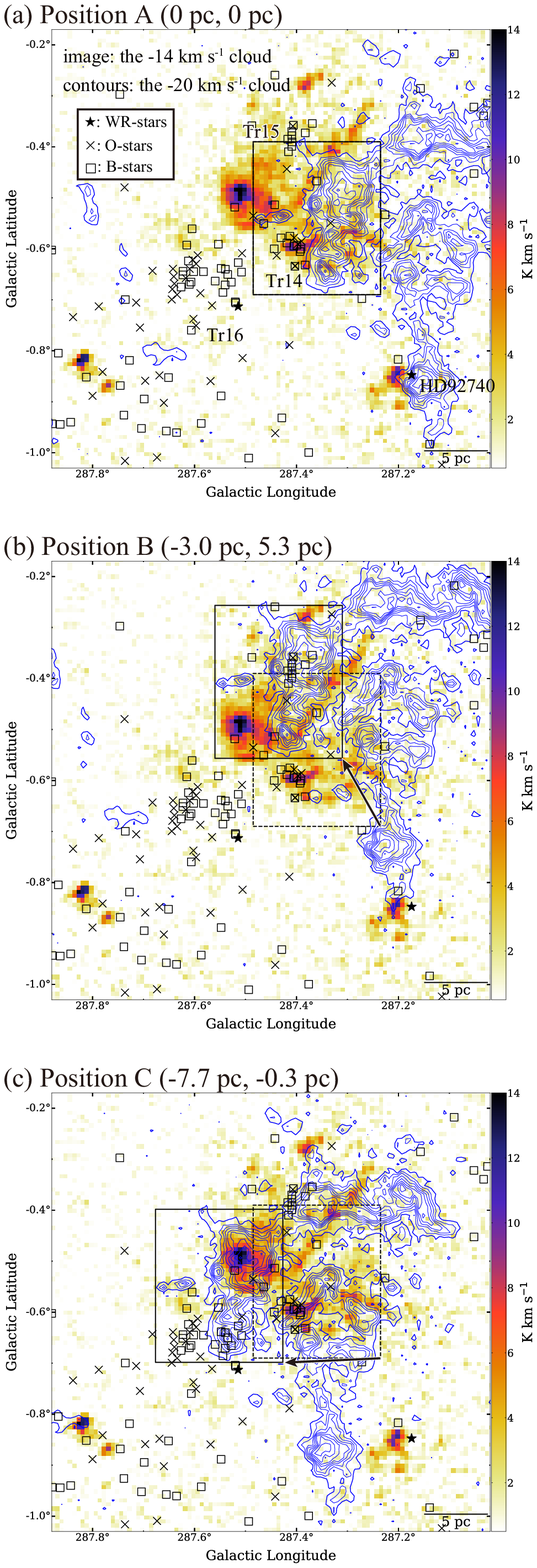}
  \end{center}
  \caption{Same as Figure\,{\ref{fig:comp}}(a). (a) The blue contours ($-20$\,km\,s$^{-1}$ cloud) are displaced by (X-axis displacement, Y-axis displacement)$=$(0\,pc, 0\,pc). (b) The blue contours are displaced by (-3.0\,pc, 5.3\,pc). (c) The blue contours are displaced by (-7.7\,pc, -0.3\,pc). }\label{fig:disp_appen}
\end{figure}

\begin{figure}[htbp]
  \begin{center}
  \includegraphics[width=12cm]{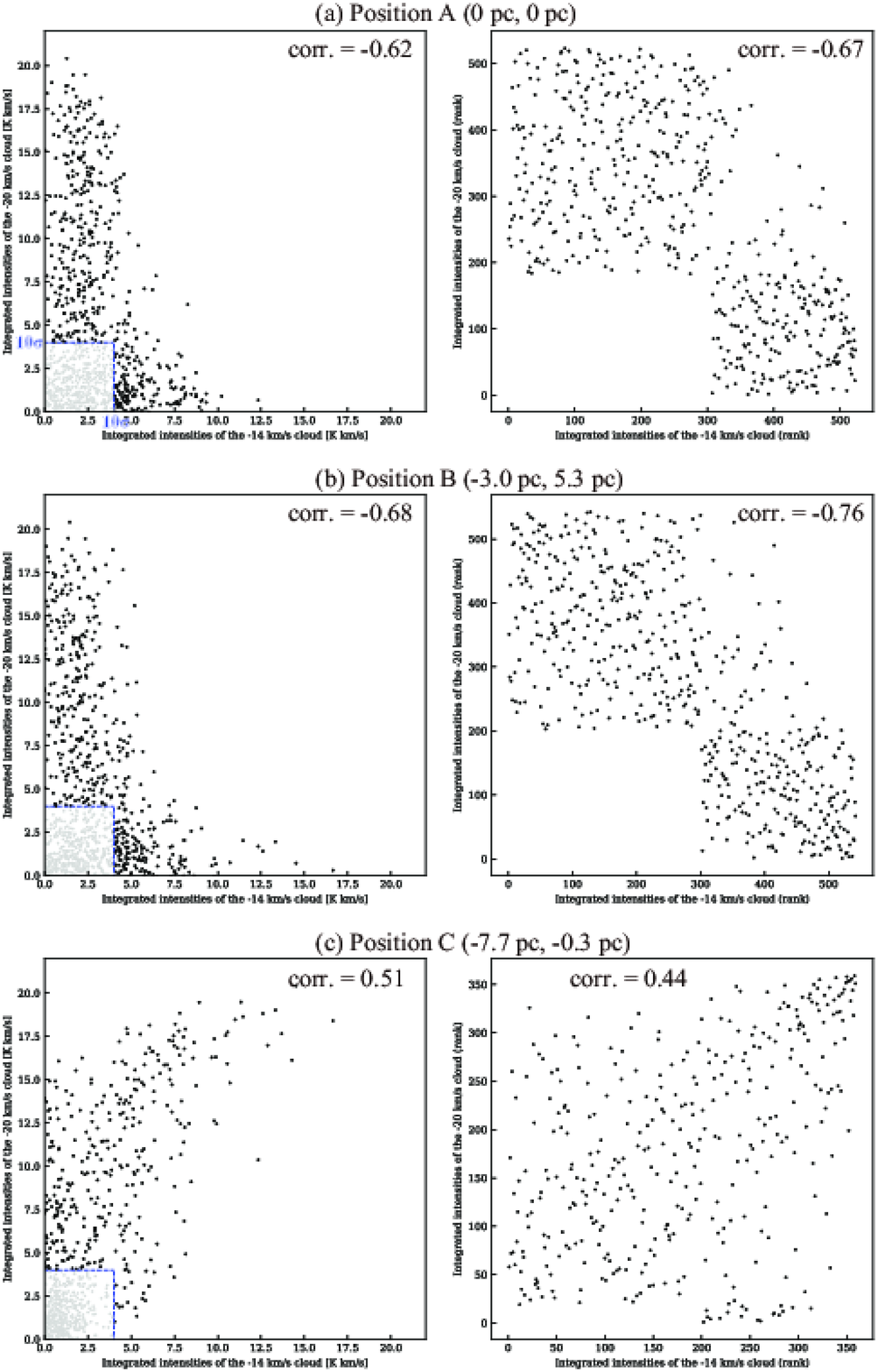}
  \end{center}
  \caption{Scatter plot of the integrated intensities of the $-14$\,km\,s$^{-1}$ cloud (X-axis) and the $-20$\,km\,s$^{-1}$ cloud (Y-axis). The left panels and the right panels show the raw integrated intensities and the rankings, respectively. }\label{fig:corr}
\end{figure}

We first defined the area of the $-20$\,km\,s$^{-1}$ cloud as shown by the black rectangle in Figure\,\ref{fig:disp_appen}(a).
Next, we calculated $r_S$ between the integrated intensities of the $-14$\,km\,s$^{-1}$ cloud (the image) and the $-20$\,km\,s$^{-1}$ cloud (the blue contours) pixel-to-pixel within the given area.
Figure\,\ref{fig:corr}(a) shows scatter plots of the integrated intensities of the two clouds. 
The left panel and the right panel show the raw integrated intensities and the rankings, respectively.
In this calculation, we removed pixels for which the values are both $<10\sigma$.
As a result, we calculate $r_S$ $-0.67$ at the original position A. 

\begin{figure}[htbp]
  \begin{center}
  \includegraphics[width=8cm]{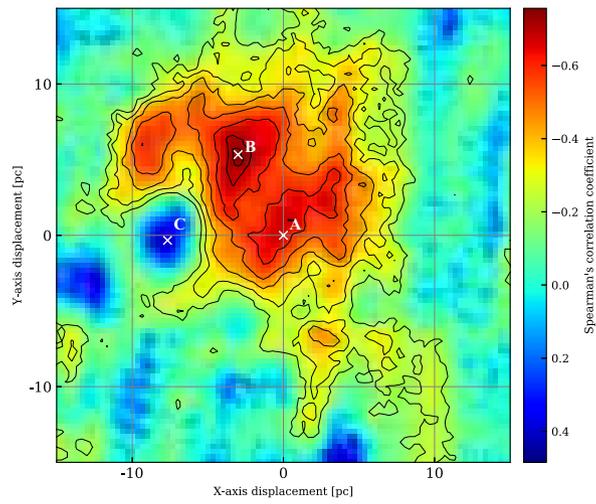}
  \end{center}
  \caption{The distribution of Spearman's correlation coefficient $r_S$ as a function of displacements along the X-axis (Galactic longitude) direction and along the Y-axis (Galactic latitude) direction. The contour levels are -0.74, -0.70, -0.60, -0.50, -0.40, -0.30, and -0.20. The white crosses labelled A, B, and C correspond to the displacements shown in Figure\,\ref{fig:disp_appen}.}\label{fig:disp_map_close}
\end{figure}

Similarly, we calculate $r_S$ for each displacement (X-axis displacement, Y-axis displacement) by a unit of one pixel.
In the present study, one pixel corresponds to $0.33$\,pc.
Figure\,\ref{fig:disp_map_close} shows a map of the calculated values of $r_S$.
The minimum occurs at the displacement (X-axis displacement, Y-axis displacement)$=$($-3.0$\,pc, $5.3$\,pc) (position B in Figure\,\ref{fig:disp_map_close}). 
There the two clouds show the most complementary distribution [Figure\,\ref{fig:disp_appen}(b)].
In contrast, at the local maximum the displacement is (X-axis displacement, Y-axis displacement)$=$($-7.7$\,pc, $-0.3$\,pc) (position C in Figure\,\ref{fig:disp_map_close}), the two clouds show similar distributions [Figure\,\ref{fig:disp_appen}(c)].

\end{document}